\documentclass[reqno,oneside,letterpaper, 11pt]{amsart}
\usepackage{latexsym}
\usepackage{mathtools}
\usepackage{amsmath}
\usepackage{amsfonts}
\usepackage{mathrsfs}
\usepackage{tensor}
\usepackage{amscd}
\usepackage{amssymb}
\usepackage{amsfonts}
\usepackage{float}
\usepackage{esint}

\usepackage{caption}
\usepackage{hyperref}
\usepackage[utf8]{inputenc}
\usepackage{color}
\allowdisplaybreaks 
\definecolor{Blue}{rgb}{0.,0.,1.}
\definecolor{Red}{rgb}{1.,0.,0.}
\definecolor{Green}{rgb}{0.,1.,0.}

\let\origmaketitle\maketitle
\def\maketitle{
  \begingroup
	\origmaketitle
  \endgroup
	}	
	
\makeatletter
\renewcommand{\author}[2][]{%
  \def\@tempa{#1}
  \ifx\@empty\authors
    \ifx\@tempa\@empty
      \gdef\shortauthors{#2}%
    \else
      \gdef\shortauthors{#1}%
    \fi
    \gdef\authors{\author{#2}}%
  \else
    \ifx\@tempa\@empty
      \g@addto@macro\shortauthors{\and#2}%
    \else
      \g@addto@macro\shortauthors{\and#1}%
    \fi
    \g@addto@macro\authors{\and\author{#2}}%
  \fi
}
\renewcommand{\address}[2][]{\g@addto@macro\authors{\address{#1}{#2}}}
\def\@setauthors{%
  \begin{center}%
    \footnotesize
    \vspace{20pt}
    \let\and\@empty
    \def\author##1{\advance\@tempcnta\@ne}%
    \def\address##1##2{\advance\@tempcntb\@ne}%
    \@tempcnta=\z@  \@tempcntb=\z@
    \authors
    \ifnum\@tempcnta>\@ne \ifnum\@tempcntb=\@ne
        \oneaddress
      \else
        \sepaddresses
      \fi
    \else
      \oneaddress
    \fi
  \end{center}%
}
\def\oneaddress{%
  \begingroup
  \let\author\@iden \let\address\@gobbletwo
  \renewcommand{\andify}{%
    \nxandlist{\unskip, }{\unskip{} and~}{\unskip, and~}}%
  \uppercasenonmath\authors
  \andify\authors
  \authors
  \endgroup
  \begingroup \let\and\relax \let\author\@gobble
  \def\address##1##2{\unskip\\[10pt] \itshape##2}%
  \authors
  \endgroup
}
\def\sepaddresses{%
  \begingroup
    \baselineskip10\p@\relax
    \def\address##1##2{ ({\itshape##2}\/)}
    \def\author##1{\def\temp{##1}\leavevmode\uppercasenonmath\temp\temp}%
    \nxandlist
      {,\\[\baselineskip]}
      {\\[\baselineskip] \textsc{\lowercase{and}}\\[\baselineskip]}
      {,\\[\baselineskip]\textsc{\lowercase{and}}\\[\baselineskip]}
      \authors 
    \authors
  \endgroup
}
\def\maketitle{\par
  \@topnum\z@
  \@setcopyright
  \thispagestyle{firstpage}%
  \uppercasenonmath\shorttitle
  \ifx\@empty\shortauthors \let\shortauthors\shorttitle
  \else
    \newcommand{\@xuppercasenonmath}[1]{\toks@\@emptytoks
      \@xp\@skipmath\@xp\@empty##1$$%
      \edef##1{\@nx\protect\@nx\@upprep\the\toks@}}%
    \@xuppercasenonmath\shortauthors
    \def\@@and{AND}
    \renewcommand{\andify}{%
      \nxandlist{\unskip, }{\unskip{ }\@@and{ }}{\unskip, \@@and{ }}}%
    \andify\shortauthors
  \fi
  \@maketitle@hook
  \begingroup
  \@maketitle
  \endgroup
  \c@footnote\z@
  \@cleartopmattertags
}
\def\@maketitle{%
  \normalfont\normalsize
  \let\@makefntext\noindent
  \@adminfootnotes
  \ifx\@empty\addresses\else \@footnotetext{\@setotheraddresses}\fi
  \global\topskip68\p@\relax
  \@settitle
  \ifx\@empty\authors \else \@setauthors \fi
  \ifx\@empty\@dedicatory
  \else
    \baselineskip26\p@
    \vtop{\centering{\footnotesize\itshape\@dedicatory\@@par}%
      \global\dimen@i\prevdepth}\prevdepth\dimen@i
  \fi
  \toks@\@xp{\shortauthors}\@temptokena\@xp{\shorttitle}%
  \edef\@tempa{\@nx\markboth{\the\toks@}{\the\@temptokena}}\@tempa
  \@setabstract
  \normalsize
  \if@titlepage
    \newpage
  \else
    \dimen@34\p@ \advance\dimen@-\baselineskip
    \vskip\dimen@\relax
  \fi
} 
\renewcommand{\thanks}[1]{%
  \ifx\@empty\thankses
    \gdef\thankses{\thanks{#1}}%
  \else
    \g@addto@macro\thankses{\endgraf\thanks{#1}}%
  \fi}
\def\@setthanks{\def\thanks##1{\noindent##1\@addpunct.}\thankses}
\renewcommand{\curraddr}[2][]{%
  \ifx\@empty\addresses
    \gdef\addresses{\curraddr{#1}{#2}}%
  \else
    \g@addto@macro\addresses{\endgraf\curraddr{#1}{#2}}%
  \fi}
\renewcommand{\email}[2][]{%
  \ifx\@empty\addresses
    \gdef\addresses{\email{#1}{#2}}%
  \else
    \g@addto@macro\addresses{\endgraf\email{#1}{#2}}%
  \fi}
\renewcommand{\urladdr}[2][]{%
  \ifx\@empty\addresses
    \gdef\addresses{\urladdr{#1}{#2}}%
  \else
    \g@addto@macro\addresses{\endgraf\urladdr{#1}{#2}}%
  \fi}
\def\@setotheraddresses{%
  \def\curraddr##1##2{\noindent
    \emph{Current address\@ifnotempty{##1}{ of ##1}}:\space
      ##2\@addpunct.}%
  \def\email##1##2{\noindent
    \emph{E-mail address\@ifnotempty{##1}{ of ##1}}:\space
      \texttt{##2}}%
  \def\urladdr##1##2{\noindent
    \emph{WWW address\@ifnotempty{##1}{ of ##1}}:\space
      \texttt{##2}}%
  \addresses
}
\let\enddoc@text\relax
\makeatother

\newcounter{smallarabics}
\newenvironment{arabicenumerate}
{\begin{list}{{\normalfont\textrm{(\arabic{smallarabics})}}}
  {\usecounter{smallarabics}\setlength{\itemindent}{0cm}
   \setlength{\leftmargin}{5ex}\setlength{\labelwidth}{4ex}
   \setlength{\topsep}{0.75\parsep}\setlength{\partopsep}{0ex}
   \setlength{\itemsep}{0ex}}}
{\end{list}}

\newcounter{smallroman}

\newcommand{\ben}{\begin{arabicenumerate}}  
\newcommand{\een}{\end{arabicenumerate}}

\def\init{\setcounter{equation}{0}}

\newtheorem{theoreme}{Theorem }[section]
\newtheorem{proposition}[theoreme]{Proposition}

\newtheorem{lemma}[theoreme]{Lemma}
\newtheorem{definition}[theoreme]{Definition}

\newtheorem{remark}[theoreme]{Remark}
\newtheorem{example}[theoreme]{Example}
\newcommand{\beq}{\begin{equation}}
\newcommand{\eeq}{\end{equation}}

\newcommand{\bex}{\begin{example}}
\newcommand{\eex}{\end{example}}
\def\bel{\begin{lemma}}
\def\eel{\end{lemma}}
\def\bet{\begin{theoreme}}
\def\eet{\end{theoreme}}
\def\bed{\begin{definition}}
\def\eed{\end{definition}}
\def\ber{\begin{remark}}
\def\eer{\end{remark}}

\def\rr{{\mathbb R}}

\def\cc{{\mathbb C}}

\def\Re{{\rm Re}}

\def\bar{\overline}

\def\reg{{\rm reg}}

\def\cinf{C^\infty}
\def\proof{
\noindent{\bf Proof.}\ \ }

\def\cV{{\mathcal V}}

\def\cD{{\mathcal D}}
\def\cU{{\mathcal U}}

\def\i{{\rm i}}
\def\qed{$\Box$\medskip}
\def\p{ \partial}
\def\12{\frac{1}{2}}

\def\Ran{{\rm Ran}}
\def\bbbone{{\mathchoice {\rm 1\mskip-4mu l} {\rm 1\mskip-4mu l}
{\rm 1\mskip-4.5mu l} {\rm 1\mskip-5mu l}}}
\def\one{\bbbone}
\def\cH{{\mathcal H}}

\def\coinf{C_0^\infty}

\def\cF{{\mathcal F}}

\def\tD{{\tilde{D}}}
\def \p{ \partial}

\def\Ran{{\rm Ran}}

\newcommand{\mat}[4]{\left(\begin{array}{cc}#1 &#2  \\ #3 &#4 \end{array}\right)}

\newcommand{\col}[2]{\left(\begin{array}{c}#1 \\#2\end{array} \right)}

\newcommand{\traa}[1]{\mskip-6mu\upharpoonright_{#1}}

\def\cE{{\mathcal E}}

\def\WF{{\rm WF}}
\makeatletter
\newcommand*{\defeq}{\mathrel{\rlap{%
                     \raisebox{0.3ex}{$\m@th\cdot$}}%
                     \raisebox{-0.3ex}{$\m@th\cdot$}}%
                     =}
\makeatother
\makeatletter
\newcommand*{\eqdef}{=\mathrel{\rlap{%
                     \raisebox{0.3ex}{$\m@th\cdot$}}%
                     \raisebox{-0.3ex}{$\m@th\cdot$}}%
                     }
\makeatother

\def\cinfb{C^{\infty}_{\rm b}}
\def\rx{{\rm x}}

\DeclareMathOperator{\Ker}{Ker}

\def\dual{\!\cdot \!}
\def\CCR{{\rm CCR}}

\def\zero{{\mskip-4mu{\rm\textit{o}}}}

\def\cN{{\mathcal N}}

\def\BT{{ BT}}

\def\calde{Calder\'{o}n }

\def\tosim{\xrightarrow{\sim}}

\def\zero{{\mskip-4mu{\rm\textit{o}}}}

\def\dual{\!\cdot \!}
\def\calde{Calder\'{o}n }
\def\rR{{\bf R}}\def\Ric{{\bf Ric}}\def\rg{{\bf g}}
\def\Riem{{\bf Riem}}\def\tD{\widetilde{D}}
\def\nab{\nabla}
\def\scal{{\bf R}}
\def\trace{{\rm tr}}
\def\dvol{\mathop{}\!d{\rm vol}}
\def\Diff{{\rm Diff}}
\def\trg{\widetilde{\rg}}
\def\rh{{\bf h}}
  \def\tV{\widetilde{V}}

\def\sc{\rm{sc}}\def\c{\rm{c}}
\def\tq{\widetilde{q}}

\begin{document}
\title[Hadamard states for  linearized gravity]{Hadamard states for  linearized gravity  on spacetimes with compact Cauchy surfaces}

\author{C. Gérard}
\address{Laboratoire de Math\'ematiques d'Orsay, Universit\'e Paris-Saclay, 91405 Orsay Cedex France}
\email{christian.gerard@math.u-psud.fr}
\keywords{linearized Einstein equations, microlocal analysis, Quantum Field Theory on curved spacetimes, Hadamard states}
\subjclass[2020]{81T20, 83C05, 58J47, 58J45, 58J32}
\thanks{\emph{Acknowledgments.} We would like to warmly thank Simone Murro and Michal Wrochna for numerous useful discussions.\\
\emph{Conflict of Interests/Competing Interests.} The author has no conflict of interests.}
\date{August 2023}

\begin{abstract}{We consider the quantization of   linearized Einstein equations. We prove the existence of Hadamard states in the harmonic gauge  on any  Einstein spacetime with  compact Cauchy  surfaces.}
\end{abstract}

\maketitle
\section{Introduction and summary}\label{sec0}\init
Linearized gravity is an example of a linear {\em gauge theory}, for which the construction of states is significantly more difficult than for ordinary matter fields.  While the structure of the classical linearized gravity  needed for its algebraic quantization is now well understood \cite{BFR, FH, BDM, HS},  the rigorous construction of physical states, ie {\em Hadamard states},  remains an important open problem.

Let us now mention several works which are related to the present one. 

The simplest example of a linear gauge theory is Maxwell equations, which were considered by 
 in  \cite{F, FP,DS}, Hadamard states being constructed in  \cite{FS}.  For linearized Yang--Mills equations around the zero solution, Hadamard states were constructed in the BRST framework in  \cite{H}. The case of linearized Yang--Mills equations around a non zero solution was considered later in  \cite{GW2}.
 
 In these models, one can use spacetime deformation arguments which are not applicable to linearized gravity. 
 
  The case of linearized gravity on asymptotically flat spacetimes was studied in  \cite{BDM} with methods drawing from earlier works  \cite{AA,DMP}, the quantization turns out however to be limited to a subspace of classical degrees of freedom due to divergences at null infinity.

More recently in \cite{GMW} the construction of Hadamard states for linearized gravity on {\em analytic spacetimes} was investigated using {\em Wick rotation}. This  consists in applying Wick rotation  in some Gaussian time coordinate with respect to a reference Cauchy surface $\Sigma$. The 
various d' Alembertian  operators are transformed into elliptic Laplacians. In general these elliptic Laplacians are only defined in some strip in imaginary Gaussian time.

One can recover a quasi-free state  for the Lorentzian theory  from {\em \calde projectors}, a well-known tool in elliptic boundary value problem. This method was first used in \cite{GW2} to construct analytic Hadamard states for scalar fields on analytic spacetimes.

States obtained from \calde projectors depend on less arbitrary choices than those constructed by pseudodifferential calculus. Therefore it is hoped that the crucial gauge invariance property will be automatically satisfied. 

However there are still a number of difficulties to obtain gauge invariant Hadamard states from \calde projectors. 

Firstly the Wick rotated operators should be not only elliptic but also invertible. To define them properly one has to impose some boundary conditions on the boundary of the strip in which they are defined.

The Dirichlet boundary conditions used in \cite{GMW} have the advantage of easily giving invertibility of the Wick rotated operators and a modified positivity property. However they  are not gauge invariant. As a consequence in \cite{GMW} the gauge invariance and positivity of the two-point functions are only obtained modulo the addition of some smooth corrections.

In this paper we reconsider the problem of existence of  Hadamard states for linearized gravity by using a different strategy. We prove the following result:
\begin{theoreme}
Let $(M, \rg)$ a globally hyperbolic spacetime with $\dim M=4$ and $\Ric= \Lambda \rg$, $\Lambda\in \rr$. Assume that $(M, \rg)$ has {\em compact Cauchy surfaces}. Then there exist Hadamard states for linearized gravity on $(M, \rg)$.
\end{theoreme}

\subsection{Description of the paper}
We now briefly recall the algebraic quantization of linearized gravity, explain the  difficulties encountered when trying to construct Hadamard states for linearized gravity, and describe the approach we use in this paper to overcome them.
\subsubsection{Linearized Einstein equations}
Let $(M, \rg)$ be a globally hyperbolic spacetime with $\dim M= 4$ solving the Einstein equations
\[
\Ric= \Lambda\rg,
\]
where $\Lambda\in \rr$ is the cosmological constant. Let $V_{k}\defeq\cc\otimes^{k}_{\rm s}T^{*}M$, $k=1,2$  be the complex bundle of symmetric $(0, k)$-tensors on $M$. We consider the two differential operators
\[
P= - \square_{2}-  I \circ d \circ \delta + 2\,\Riem_{\rg}, \ K= I\circ d,
\]
where 
\ben\setlength{\itemindent}{0,5cm}
\item[-] $\square_{2}$ is the  d'Alember\-tian, $(\square_{2} u)_{ab} =  \nab^{c}\nab_{c}u_{ab}$, 

\item[-] $I$ is the trace reversal $(Iu)_{ab}= u_{ab}- \frac{1}{2} \trace_{\rg}(u) \rg_{ab}$,
\item[-] $d$ the symmetric differential $(d w)_{a b}= \nab_{(a}u_{b)}$,
\item[-]  $\delta$ is the formal adjoint of $d$,  $(\delta u)_{a}= -2\nab^{c}u_{ca}$,
\item[-] $\Riem_{\rg}$ is the Ricci operator $(\Riem_{\rg}u)_{ab}= \rR\indices{_{a}^{cd}_{b}}u_{cd}$.
\een \smallskip
The {\em linearized Einstein equations} around $\rg$ are 
\begin{equation}
\label{e0.1}
Pu=0,
\end{equation}
where $u$ is a (symmetric) $(0,2)$-tensor. 
The identity $P\circ K=0$ implies that $\Ker_{\rm sc}P$ is invariant under 
 {\em linearized gauge transformations} given by $u\mapsto u+ K w$, where $w$ is a $(0, 1)$-tensor. Therefore the natural 'on-shell' phase space is  the quotient space:
\[
\dfrac{\Ker_{\sc}P}{\Ran_{\sc}K}.
\]
Here  and below the subscripts $\sc$ resp.  $\c$ refer to 'space compact', resp. 'compact' for example $\Ker_{\sc}P$ is the space of (smooth) space compact solutions of \eqref{e0.1}.

\subsubsection{The phase space for linearized gravity}
The operator $P$ is not hyperbolic, hence does not have advanced/retarded propagators. To equip the phase space with a Hermitian structure, it is necessary to add a {\em subsidiary gauge condition}.  We follow here the nice exposition in \cite{HS}. In this paper we will use  the {\em de Donder} or {\em harmonic gauge}:
\[
K^{\star}u=0,
\]
where $K^{\star}= \delta$ is the adjoint of $K$ for a Hermitian form $(\cdot| \cdot)_{I, V_{2}}$ involving $I$, see \ref{sec1.5.1}, for which $P$ is formally selfadjoint. The  quotient space $\dfrac{\Ker_{\sc}P}{\Ran_{\sc}K}$ is then isomorphic to
\[
\dfrac{\Ker_{\rm sc}D_{2}\cap \Ker_{\rm sc}K^{\star}}{K\Ker_{\rm sc}D_{1}},
\]
where
\[
\begin{array}{l}
D_{1}= K\circ K^{\star}= - \Box_{1}-\Lambda, \\[2mm]
D_{2}= P+ K\circ K^{\star}=- \Box_{2}+ 2\Riem_{\rg}
\end{array}
\]
are hyperbolic operators acting respectively on $(0,1)$- and $(0, 2)$-tensors. Since $D_{2}$ is hyperbolic, it admits advanced/retarded propagators $G_{2\,{\rm ret/adv}}$. On can then introduce   the 'off shell' phase space:
\[
\cV_{P}= \dfrac{\Ker_{\c}K^{\star}}{\Ran_{\c}P}
\]
equipped with the Hermitian form
\[
\bar{[u]}\dual Q_{P}[u]\defeq  \bar{u}\dual Q_{I, 2}u, 
\]
where 
\[
\bar{u}\dual Q_{I, 2}u= \i (u|I G_{2} u)_{V_{2}}, 
\]
and $G_{2}= G_{2\,{\rm ret}}- G_{2\,{\rm adv}}$ is the commutator function for $D_{2}$.

The algebraic quantization of linearized gravity simply consists  in constructing the  $\CCR$ $*$-algebra $\CCR(\cV_{P}, Q_{P})$. 

Note that other Hermitian spaces, isomorphic to $(\cV_{P}, Q_{P})$ are useful. In this paper, after fixing a reference Cauchy surface $\Sigma$,  we will rely on the Hermitian space  of  Cauchy data
$(\dfrac{\Ker_{\c} K_{\Sigma}^{\dagger}}{\Ran_{\c} K_\Sigma}, q_{I, 2})$, where $K_{\Sigma}, K_{\Sigma}^{\dag}$ are Cauchy surface analogs of $K, K^{\star}$, see \ref{sec1.6.2}. 

\subsubsection{Hadamard states}
A quasi-free state on  $\CCR(\cV_{P}, Q_{P})$  is defined by a pair of Hermitian forms $\Lambda^{\pm}_{P}$ on $\cV_{P}$ called {\em covariances} such that 
 \beq\label{e0.2}
\begin{array}{rl}
i)&\Lambda^{\pm}_{P}= \Lambda_{P}^{\pm*}, \ \Lambda_{P}^{\pm}\geq 0,\\[2mm]
ii)&\Lambda_{P}^{+}- \Lambda_{P}^{-}= Q_{P}.
\end{array}
\eeq
We will be interested in covariances obtained from 
a pair of sesquilinear forms on $\coinf(M; V_{2})$ by
\beq\label{e0.3}
\bar{[u]}\dual \Lambda^{\pm}_{P}[u]\defeq \bar{u}\dual \Lambda_{2}^{\pm}u,
\eeq
The forms  $\Lambda_{2}^{\pm}$ have to satisfy a number of conditions. 

Firstly  they should 'pass to quotient' i.e. \eqref{e0.3} should be meaningful.  This leads to the conditions:
\[
\begin{array}{rl}
(1)&D_{2}^{*}\circ \Lambda_{2}^{\pm}= \Lambda_{2}^{\pm}\circ D_{2}=0,\\[2mm]
(2)& \Lambda_{2}^{\pm}=0 \hbox{ on  } \Ker_{\c}K^{\star}\times \Ran_{\c}K.
\end{array}
\]
Condition (1) corresponds to  the  {\em field equations}, familiar from quantization of matter fields, while condition (2), specific to gauge fields, is the {\em gauge invariance}.

The next two conditions are  
\[
\begin{array}{rl}
(3)& \Lambda_{2}^{\pm}= \Lambda_{2}^{\pm*}, \Lambda_{2}^{\pm}\geq 0, \hbox{ on } \Ker_{\c}K^{\star},\\[2mm]
(4)& \Lambda_{2}^{+}- \Lambda_{2}^{-}=Q_{I, 2},
\end{array}
\]
and correspond to \eqref{e0.2}. Condition (3) is the {\em positivity}, while condition (4) corresponds to the {\em CCR}.

The last condition is  the {\em Hadamard condition}, which singles out Hadamard states, considered as the physically meaningful states on $\CCR(\cV_{P}, Q_{P})$. Denoting by $\lambda_{2}^{\pm}\in\cD'(M\times M; L(V_{2}))$ the distributional kernels of $\Lambda_{2}^{\pm}$, one requires that
\[
(5) \  \WF(\lambda_{2}^{\pm})'\subset \cN^{\pm}\times \cN^{\pm},
\]
where $\cN^{\pm}$ are the two connected components of the characteristic manifold $\cN= \{(x, \xi)\in T^{*}M\setminus\zero: \xi\dual \rg^{-1}(x)\xi= 0\}$.

Conditions (1), (4) and (5) are rather easy to satisfy. In fact the construction of Hadamard states for scalar fields via pseudodifferential calculus initiated in \cite{junker, junkererratum} and further developped in \cite{GW1, GOW} can be adapted to the tensor case and produces a wealth of covariances satisfying (1), (4) and (5), see \cite[Sect. 5]{GMW}.

Condition (3) (positivity) is much more delicate, because $D_{2}$ is selfadjoint only for a {\em non-positive} Hermitian form. This difficulty is  at the origin of use of Krein spaces ('Hilbert spaces' with  a non-positive scalar product) appearing in the Gupta-Bleuler approach in QED.

If true, condition (3) will in general only be satisfied on the subspace $\Ker_{\c}K^{\star}$.

Condition (2) (gauge invariance) is also very difficult to impose, because it has to be satisfied exactly, not only modulo smoothing errors.

\subsubsection{The approach in this paper}

In this paper we circumvent the difficulties  with conditions (2) and (3) by relying on {\em full gauge fixing}.

Working  with the Cauchy surface phase space $\dfrac{\Ker_{\c}K_{\Sigma}^{\dag}}{\Ran_{\c}K_{\Sigma}}$, we start with a pair $\lambda_{2\Sigma}^{\pm}$ of  Cauchy surface covariances, see \ref{sec1.7.3}, whose associated $\Lambda_{2}^{\pm}$ will satisfy (1), (4) and (5). 

We next try to  find a convenient supplementary space $E$ of $\Ran_{\c}K_{\Sigma}$ in $\Ker_{\c}K_{\Sigma}^{\dag}$. We can then identify the canonical phase space $\dfrac{\Ker_{\c}K_{\Sigma}^{\dag}}{\Ran_{\c}K_{\Sigma}}$ with $E$ using the associated projection
$\pi: \Ker_{\c}K_{\Sigma}^{\dag}\to E$.

 The {\em modified covariances} $\pi^{*}\circ \lambda_{2\Sigma}^{\pm}\circ \pi$ will then automatically satisfy the gauge invariance condition. 

The supplementary space $E$ has however to be chosen appropriately. First $\lambda_{2\Sigma}^{\pm}$ should be {\em positive} on $E$ if the positivity condition (3) is to be satisfied by the modified covariances.  Second $E$ has also to be adapted so that the projection $\pi$ does not spoil the microlocal Hadamard condition (5).

We select the space $E$ using a microlocal version of the {\em synchronous gauge condition}, see \ref{sec10.1.2}. The fact that $E$ is supplementary to $\Ran_{\c}K_{\Sigma}$ is equivalent  to the solvability of some  {\em elliptic system} of equations on Cauchy data. 

If the  system is uniquely solvable, (the so called {\em regular case}, see Subsect. \ref{sec10.4}), then the existence of a Hadamard state for linearized gravity follows rather easily.

If it is not uniquely solvable (the so called {\em singular case}, see Subsect. \ref{sec10.5})  and if the Cauchy surface $\Sigma$ is {\em compact}, 
then by Fredholm theory it still has finite dimensional kernel and cokernel. We can  further alter the modified covariances by some finite rank and smoothing operators to obtain a pair of Hadamard covariances.

\subsection{Notation}\label{sec0.2}
We now collect various notations used throughout the paper.
\subsubsection{Isomorphisms of vector spaces}
If $E, F$ are vector spaces and $A\in L(E, F)$ we write $A: E\tosim F$ if $A$ is an isomorphism. If $E, F$ are topological vector spaces, we use the same notation if $A$  is a homeomorphism.
\subsubsection{Sesquilinear forms}
If $E$ is a complex vector space, its antidual is denoted by $E^{*}$. 
A sesquilinear form $A$ on $E$ is an element of $L(E, E^{*})$ and its action on elements of $E$ is denoted by $\bar{u}\dual Av$. 
\subsubsection{Projections}
If $F\subset E$ are two vector spaces we say that $\pi: E\to F$ is a projection if  $\pi^{2}= \pi$ and  $\Ran \pi= F$.

\subsubsection{Operators on quotient spaces}\label{sec0.2.1}

Let $F_i\subset E_i$, $i=1,2$ be vector spaces and let $A\in L(E_1, E_2)$. Then the induced map
\[
[A]\in L( E_1/F_1, E_2/F_2),
\]
 is
\begin{equation}
\label{e0.01}
\begin{array}{rl}
1)&\hbox{well-defined if }A E_1\subset E_2\hbox{ and }A F_1\subset F_2,\\[2mm]
2)&\hbox{injective iff }A^{-1}F_2=F_1,\\[2mm]
3)&\hbox{surjective iff }E_2=A E_1+F_2. 
\end{array}
\end{equation}

\subsubsection{Sesquilinear forms on quotients}\label{sec0.2.2} 
Let now $E\subset F$ be vector spaces and let $C\in L( E, E^*)$. We denote by $F^\circ\subset E^*$ the annihilator of $F$. Then the induced map
\[
[C]\in L( E/F,(E/F)^*),
\]
 is
\begin{equation}
\label{e0.02}
\begin{array}{rl}
1)&\hbox{well-defined if }CE\subset F^{\circ}, F\subset\Ker\, C,\\[2mm]
2)&\hbox{non-degenerate iff }F=\Ker\,C. 
\end{array}
\end{equation}
If $C$ is hermitian or anti-hermitian then the condition $F\subset\Ker\, C$ implies the other one $CE\subset F^{\circ}$ (and vice versa).

\subsubsection{Sections of vector bundles}\label{sec01.1.1}
Let  $V\xrightarrow{\pi}M$ be  a finite rank complex vector bundle  over a smooth manifold $M$.  

- If $\Sigma\subset M$ is a smooth manifold we denote by $V|_{\Sigma}\xrightarrow{\pi}\Sigma$ the restriction of $V$ to $\Sigma$.

- We denote by $\cinf(M; V)$, resp.~$\coinf(M; V)$ the space of smooth, resp.~ compactly supported smooth sections of $V$.  
 
 -We denote by $\cD'(M; V)$, resp.~$\cE'(M; V)$ the space of distributional, resp.~ compactly supported distributional  sections of $V$.

We use the same notations if $V$ is a finite dimensional vector space, i.e.~we write simply $V$ instead of the trivial vector bundle $M\times V$.

\subsubsection{Globally hyperbolic spacetimes}
We use the convention  $(-,+,\dots,+)$ for the Lo\-rent\-zian signature. 

- If $(M, \rg)$ is a spacetime,we denote by $J_{\pm}(K)$ the future/past causal shadow of  $K\subset M$.

- If $M$ is a globally hyperbolic spacetime  we   denote by $\cinf_{\rm sc}(M;V)$ the space of space-compact sections, i.e.~sections in $\cinf(M;V)$ with compactly supported restriction to a Cauchy surface.

\subsubsection{Distributional kernels and wavefront sets}

 -If $u\in \cD'(M; V)$ we denote by $\WF (u)\subset T^{*}M\setminus\zero$ its {\em wavefront set}, which is invariantly defined using local trivializations of $V$. 

-If $V_{i}\xrightarrow{\pi}M_{i}$ are two vector bundles as above and $A: \coinf(M_{1}; V_{1})\to \cD'(M_{2}; V_{2})$ is linear continuous, then $A$ admits a distributional kernel, still denoted by $A\in \cD'(M_{2}\times M_{1}; V_{2}\boxtimes V_{1})$.  

- We denote by $\WF(A)'\subset (T^{*}M_{2}\times T^{*}M_{1})\setminus \zero$ its {\em primed} wavefront set, defined by 
  \[
  \Gamma'= \{((x_{2}, \xi_{2}), (x_{1}, - \xi_{1})): ((x_{2}, \xi_{2}), (x_{1}, \xi_{1}))\in \Gamma\}\hbox{ for }\Gamma\subset T^{*}M_{2}\times T^{*}M_{1}.
  \]

\section{Linearized gravity}\label{sec1}\init
In this section we review the quantization of linearized gravity, following \cite{HS}. We also introduce  the useful phase spaces of Cauchy data, following \cite{GW2}.
\subsection{Notation and background} \label{sec1.1}
 We start by fixing notation. Let $(M,\rg)$ be a $4$-dimensional  Lorentzian manifold.   
\subsubsection{Convention for the Riemann tensor}\label{sec1.1.1} 
We use the same convention as in e.g.~\cite{R,FH,BDM} for the sign of the Riemann tensor i.e.
\[
(\nab_{a}\nab_{b}- \nab_{b}\nab_{a}) u_{\c}= \rR\indices{_{abc}^{d}}u_{d}
\]
on $(0,1)$-tensors. The Ricci tensor  is $\Ric_{ab}=\rR\indices{_{acb}^{c}}= \rR\indices{^{c}_{acb}}$,
and the scalar curvature $\scal= \rg^{ab}\Ric_{ab}$. The Einstein equations with cosmological constant $\Lambda$, i.e.
$
\Ric -\12\rg\scal +\Lambda \rg =0
$,
are equivalent to 
\beq\label{eq:einstein}
\Ric =\Lambda \rg.
\eeq 
 We will say that $(M, \rg)$ is Einstein if \eqref{eq:einstein} is satisfied.
\subsubsection{Hermitian forms on tensors}\label{sec1.1.2}
 We denote by 
\[
V_{k}\defeq\cc\otimes^{k}_{\rm s}T^{*}M
\]
 the complex bundle of symmetric $(0,k)$-tensors.  We will  only need  the cases $k=0, 1,2$.
$V_{k}$ is  equipped with the non-degenerate Hermitian form
\begin{equation}
\label{eq:uvk}
(u| u)_{V_{k}}\defeq k! \bar{u}\dual (\rg^{\otimes k})^{-1}u.
\end{equation}
 In abstract index notation,
\[
 (u|u)_{V_{k}}= k! \,\rg^{a_{1}b_{1}}\cdots\rg^{a_{k}b_{k}}\bar{u}_{a_{1}\dots a_{k}} u_{b_{1}\dots b_{k}}.
\]
 For example for $k=2$ we have
\beq\label{e0.00}
(u|u)_{V_{2}}= 2\trace(u^{*}\rg^{-1}u\rg^{-1}).
\eeq
The $k!$ normalization differs from the most common convention, it has however the advantage that various expressions involving adjoints look more symmetric.

For $U\subset M$ open,  the Hermitian form \eqref{eq:uvk} on fibers induces a Hermitian form
\begin{equation}
\label{uvkU}
(u|v)_{V_{k}(U)}=  \int_{U}(u(x)| v(x))_{V_{k}}\dvol_{\rg}, \quad u, v\in \coinf(U;V_{k}).
\end{equation} 
The adjoint of  $A: \cinf(M; V_{k})\to \cinf(M; V_{l})$ for those Hermitian forms will be denoted by $A^{*}$.

If $\Sigma\subset M$ is a Cauchy surface, we set
\[
(u|v)_{V_{k}(\Sigma)}=  \int_{\Sigma}(u(x)| v(x))_{V_{k}}\dvol_{\rh}, \quad u, v\in \coinf(\Sigma;V_{k}),
\]
where $\dvol_{\rh}$ is the induced density on $\Sigma$.
\subsubsection{Decomposition of tensors}\label{sec1.1.3}
\def\Sig{\Sigma}
Let us assume that $M= I\times \Sigma$ where $I\subset \rr$ is an open interval,   $\Sigma$ a smooth manifold  with variables $(t, \rx)$ and
\[
\rg= - dt^{2}+ \rh(t, \rx)d\rx^{2},
\]
where $ \rh\in \cinf(M,  \otimes^{2}_{s}T^{*}\Sigma)$ is a smooth $t$-dependent Riemannian metric on $\Sigma$.  We set 
\[
V_{k\Sigma}= \cc\otimes_{\rm s}^{k}T^{*}\Sigma.
\]
\subsubsection{ Decomposition of $(0,1)$-tensors}\label{sec1.1.4}
 We identify  \beq\label{etiti.0}
 \begin{array}{l}
 \cinf(M; V_{1})\tosim\cinf(I; \cinf(\Sigma; V_{0\Sigma}))\oplus \cinf(I; \cinf(\Sigma; V_{1\Sigma}))\hbox{ by}\\[2mm]
  w\mapsto (w_{t}, w_{\Sig}), \\[2mm]
   w \eqdef w_{t}dt+ w_{\Sig}.
  \end{array}
 \eeq
  The scalar product $(\cdot| \cdot)_{V_{1}}$ reads then
 \[
 (w| w)_{V_{1}}= - | w_{t}|^{2}+ (w_{\Sig}| w_{\Sig})_{V_{1\Sigma}}=  - | w_{t}|^{2}+ (w_{\Sig}|\rh^{-1} w_{\Sig}).
 \]
\subsubsection{Decomposition of $(0,2)$-tensors}\label{sec1.1.5}
Similarly we identify  \beq\label{etiti.-1}
 \begin{array}{l}
 \cinf(M; V_{2})\tosim\cinf(I; \cinf(\Sigma; V_{0\Sigma}))\oplus \cinf(I; \cinf(\Sigma; V_{2\Sigma}))\oplus\cinf(I; \cinf(\Sigma; V_{2\Sigma}))\hbox{ by}\\[2mm]
  u\mapsto (u_{tt}, u_{t\Sig},u_{\Sig\Sig}), \\[2mm]
u\eqdef u_{tt}dt\otimes dt+ u_{t\Sig}\otimes dt+ dt\otimes u_{t\Sig}+ u_{\Sig\Sig}.
  \end{array}
 \eeq

The scalar product  $(\cdot| \cdot)_{V_{2}}$  reads:
\beq\label{e2.9b}
(u|u)_{V_{2}}= 2|u_{tt}|^{2}-  4(u_{t\Sig}| u_{t\Sig})_{V_{1\Sigma}} +  (u_{\Sig\Sig}| u_{\Sig\Sig})_{V_{2\Sigma}}.
\eeq

\subsection{The differential and its adjoint}\label{sec1.2}
Let  
\[
d: \begin{array}{l}
\cinf(M; V_{k})\to \cinf(M; V_{k+1})\\[1mm]
(d u)_{a_{1} \dots, a_{k+1}}= \nab_{(a_{1}}u_{a_{2}\dots, a_{k+1})},
\end{array}
\]
where $u_{(a_{1} \dots a_{k})}$ is the symmetrization of $u_{a_{1}\dots a_{k}}$,
and
\[
\delta: \begin{array}{l}
\cinf(M; V_{k})\to\cinf(M; V_{k-1})\\[1mm]
(\delta u)_{a_{1}, \dots, a_{k-1}}= -k\nab^{a}u_{aa_{1}\dots a_{k-1}}.
\end{array}
\]
With these conventions, we have $d^*=\delta$ w.r.t.~the Hermitian form \eqref{uvkU}.

\subsection{Operators on tensors}\label{sec1.3}
\subsubsection{Trace reversal}\label{sec1.3.1}
The operator of \emph{trace reversal} $I$
 is given by
 \[
I\defeq \one - \frac{1}{4}|\rg)(\rg|,
\]
where
\[
(\rg| : u_{2}\mapsto (\rg| u_{2})_{V_{2}}, \ |\rg): u_{0}\mapsto u_{0}\rg,
\]
i.e.~$I$ is the orthogonal symmetry w.r.t.~the line $\cc \rg$. Equivalently
\[
(Iu)_{ab}= u_{ab}- \frac{1}{2} \trace_{\rg}(u) \rg_{ab}, \quad \trace_{\rg}(u)\defeq \rg^{ab}u_{ab}= \12 (\rg| u)_{V_{2}}. 
\]
It satisfies 
\beq\label{idiot}
I^{2}= \one, \quad I= I^{*}\hbox{ on }\cinf(M; V_{2}).
\eeq
\subsubsection{Ricci operator}\label{sec1.3.2}
The \emph{Ricci operator} is
\[
\Riem_{\rg}(u)_{ab}\defeq \rR\indices{_{a}^{cd}_{b}}u_{cd}= \rR\indices{^{c}_{ab}^{d}}u_{cd}, \ \ u\in\cinf(M; V_{2}).
\]
The fact that $\Riem_{\rg}$ preserves symmetric $(0,2)$-tensors follows from the symmetries of the Riemann tensor. 
\begin{lemma}\label{lemma2.12} The Ricci operator satisfies:
 \beq\label{e2.5}
\begin{array}{rl}
i)&\Riem_{\rg} \rg=-\Ric,\\[2mm]
ii)&\Riem_{\rg} \circ I=  I\circ \Riem_{\rg},\hbox{ if }g\hbox{ is Einstein},\\[2mm]
iii)&\Riem_{\rg}= \Riem_{\rg}^{*}.
\end{array}
\eeq
\end{lemma}

\subsection{Lichnerowicz operators}\label{sec1.4}
Let  $-\square_{i}$  be the rough d'Alembertian acting on  sections of $V_{k}$:
\[
-\square_{i} u_{i}= - \rg^{ab}\nabla^{2}_{e^{a}, e^{b}}u_{i}, 
\]
where $(e_{a})_{0\leq a\leq d}$ is a local frame.
The {\em Lichnerowicz operators}  \cite{L} acting on sections of $V_{k}$ are defined by:
 \beq\label{e0.3bip}
 \begin{array}{l}
 D_{0,L}= -\square_{0},\\[2mm]
 D_{1, L}= - \square_{1} + \Ric\circ \rg^{-1} ,\\[2mm]
D_{2, L}= - \square_{2}+ \Ric\circ \rg^{-1} \circ\cdot+  \cdot\circ \rg^{-1}\circ \Ric + 2 \Riem_{\rg} .
 \end{array}
 \eeq
One has 
\[
D_{i, L}= D_{i, L}^{*}.
\]
The proofs of the following facts can be found for example in \cite{B}.
\begin{proposition}\label{prop0.-1}
 If $(M, \rg)$ is Einstein then:
 \[
 \begin{array}{l}
  D_{i+1, L}\circ d= d\circ D_{i, L}, \ \delta \circ D_{i+1, L}= D_{i, L}\circ \delta, \\[2mm]
  (\rg| \circ D_{2, L}= D_{0, L}\circ (\rg|, \ D_{2, L}\circ |\rg) = |\rg) \circ D_{0, L}.
  \end{array}
 \]
 \end{proposition}

\subsection{Linearized gravity as a gauge theory} \label{sec1.5}
In this subsection we follow \cite{FH, HS}. Let $(M,\rg)$ be a globally hyperbolic spacetime of dimension $4$. We assume that $(M,\rg)$ is Einstein.
Let us introduce the differential operators
\beq\label{e2.4bb}
\begin{array}{rl}
P&\defeq - \square_{2}-  I \circ d \circ \delta + 2\,\Riem_{\rg},\\[2mm]
K&\defeq  I\circ d.
\end{array}
\eeq
$Pu=0$ are  the \emph{linearized Einstein equations}.  The condition $K^\star u=0$, where $K^{\star}$ is defined below,  is the linearized \emph{de Donder} or \emph{harmonic gauge}.

\subsubsection{Physical Hermitian form}\label{sec1.5.1}
We consider $V_{k}$, $k=0,1,2$ as Hermitian bundles, where the Hermitian forms on fibers is now
\beq\label{def-physical-scalar-product}
(u| u)_{I,V_{k}} \defeq (u|u)_{V_{k}},\ k= 0, 1, \  (u| u)_{I,V_{2}} \defeq (u| I u)_{V_{2}}. 
\eeq
The corresponding Hermitian form on smooth sections of $V_k$, $k=1,2$ is 
\begin{equation}\label{eq:hform}
(u|u)_{I,V_{k}(U)}=  \int_{U}(u(x)| u(x))_{I,V_{k}}\dvol_\rg, \quad u, v\in \coinf(U;V_k).
\end{equation}

We denote by $A^{\star}$ the corresponding formal adjoint of $A$ for $(\cdot| \cdot)_{I, V_{k}(M)}$ to distinguish it from the formal adjoint $A^*$ for   $(\cdot|\cdot)_{V_{k}(M)}$.  The two are related as follows:
\beq\label{e2.7}
\begin{array}{rl}
A^{\star}= I A^{*}I \hbox{ if }A: \cinf(M;V_{2})\to \cinf(M;V_{2}),\\[2mm]
A^{\star}= A^{*}I\hbox{ if }A: \cinf(M;V_{k})\to \cinf(M;V_{2}), \ k= 0, 1\\[2mm]
A^{\star}= IA^{*}\hbox{ if }A: \cinf(M;V_{2})\to \cinf(M;V_{k}),\ k= 0, 1\\[2mm]
A^{\star}= A^{*}\hbox{ if }A: \cinf(M;V_{i})\to \cinf(M;V_{j})\ i, j\neq 2.
\end{array}
\eeq 
In particular,
\beq\label{eq:Kstar}
K^{\star}= K^{*}\circ I=\delta \circ I \circ I=\delta.
\eeq 
\subsubsection{Operators in linearized gravity}\label{sec1.5.1b}
Let us set:
\[
D_{k}\defeq D_{k, L}- 2\Lambda, \ k=0, 1,2.
\]
Then 
\beq\label{e0.6}
\begin{array}{l}
 K^\star K= D_{1} =-\square_{1}-\Lambda,\\[2mm]
P+KK^\star= D_2=-\square_{2} +2\Riem_{\rg}.
\end{array}
\eeq
The operator $D_{0}$ is useful in connection with the {\em traceless gauge}. Note that 
\[
P= P^{\star}, \ D_{2}= D_{2}^{*}= D_{2}^{\star}, \ [I, D_{2}]=0.
\]
The operators $D_{k}$ are Green hyperbolic  and hence admit unique {\em retarded/advanced} inverses $G_{k{\rm ret/adv}}$. The {\em causal propagators} are
\[
G_{k}\defeq G_{k \, {\rm ret}}- G_{k\,{\rm adv}},
\]
and satisfy $G_{k}^{\star}=G_{k}^{*}= - G_{k}$.

\subsection{Cauchy problem}\label{sec1.6b}
Let $\Sigma\subset M$ a smooth space-like Cauchy surface.  
For $k=0, 1,2$  we set
\[
\varrho_{k}u= \col{u\traa{\Sigma}}{\i^{-1}\nabla_{\nu}u\traa{\Sigma}}= \col{f_{0}}{f_{1}}, \ u\in \cinf_{\sc}(M; V_{k}),
\]
where $\nu$ is the future directed unit normal to $\Sigma$. 

We denote by $U_{k}$ the operator solving the Cauchy problem for $D_{k}$ i.e.
\beq\label{e00.3}
\begin{cases}
D_{k}U_{k}= 0, \\
\varrho_{k}U_{k}= \one.
\end{cases}
\eeq
\subsubsection{Conserved charges}\label{sec1.5.3}
There exist  a unique Hermitian form $q_{k}: \coinf(\Sigma; V_{k}\otimes \cc^{2})\to \coinf(\Sigma; V_{k}\otimes \cc^{2})^{*}$ called the {\em charge} of $D_{k}$,  such that 
\[
(\phi_{k}|\i G_{k}\phi_{k})_{V_{k}(M)}= \overline{\varrho_{k}u_{k}}\dual q_{k}\varrho_{k}u_{k}
\]
for $\phi_{k}\in \coinf(M; V_{k})$ and $u_{k}= G_{k}\phi_{k}\in \Ker_{\sc}D_{k}$.
One can compute $q_{k}$ using the identity
\[
\begin{array}{rl}
&(u_{k}| D_{k}v_{k})_{V_{k}(J_{\pm}(\Sigma))}- (D_{k}u_{k}| v_{k})_{V_{k}(J_{\pm}(\Sigma))}\\[2mm]
=& \pm \i^{-1}\overline{\varrho_{k}u_{k}}\dual q_{k}\varrho_{k}u_{k}, \ u_{k}, v_{k}\in \coinf(M; V_{k}).
\end{array}
\]
\subsubsection{Operators on  Cauchy data and physical charge}\label{sec1.6.1}
We follow here \cite[Subsect. 2.4]{GW2}.

To the operator $K$ we associate an operator $K_{\Sigma}$ acting on Cauchy data by setting
\beq\label{eq:relc}
K_{\Sigma}\defeq \varrho_2 K U_1.
\eeq
Similarly since $[I, D_{2}]=0$ we can define
\beq\label{eq:reld}
I_{\Sigma}\defeq \varrho_{2}I U_{2}= I\otimes \cc^{2}.
\eeq
We obtain that  $q_{2}I_{\Sigma}= I_{\Sigma}^{*}q_{2}$ and 
as in \ref{sec1.5.1} we define the Hermitian form
\[
 q_{I, 2}\defeq q_{2}\circ  I_{\Sigma},
\]
called the {\em physical charge} for $D_{2}$. 

We denote by  $K^{\dagger}_{\Sigma}$  the adjoint of $K_{\Sigma}$ for the Hermitian forms $q_{1}, q_{I, 2}$ i.e.
\beq\label{eq:defkdag}
\overline{K_{\Sigma}^{\dag}f}_{2}\dual  q_{1}f_{1}= \overline{f}_{2}\dual q_{I, 2} K_{\Sigma}f_{1}, \ f_{k}\in \coinf(\Sigma, V_{k}\otimes \cc^{2}).
\eeq
We have:
\[
K_{\Sigma}^{\dag}= \varrho_{1}K^{\star}U_{2}.
\]

\begin{lemma}\label{lem:cauchyrel}We have:
\ben
\item\label{cauchyrelit1} $K \circ U_1 = U_2 \circ  K_\Sigma$ , $K^\star \circ  U_2=U_1 \circ  K^{\dagger}_{\Sigma}$;
\item\label{cauchyrelit0} $\varrho_2  \circ K =K_{\Sigma} \circ \varrho_1$ on $\Ker_{\sc} D_1$,  $\varrho_1 \circ  K^\star=K^{\dagger}_{\Sigma} \circ \varrho_2$ on $\Ker_{\sc} D_2$;
\item\label{cauchyrelit4} $K^{\dagger}_{\Sigma}  \circ K_{\Sigma}=0$.
\een
\end{lemma}

\subsection{Phase spaces}\label{sec1.6}
\begin{proposition}\label{prop:Gpasses} The maps
\[
\begin{array}{l}
[G_2]:\dfrac{\Ker_{\c} K^\star}{\Ran_{\c} P}\longrightarrow \dfrac{\Ker_{\sc} P}{\Ran_{\sc} K},\\[3mm]
[Id]:\dfrac{\Ker_{\sc} D_2\cap\Ker_{\sc} K^\star}{K\Ker_{\sc}D_{1}}     \longrightarrow \dfrac{\Ker_{\sc} P}{\Ran_{\sc} K},
\end{array}
\]
are  well defined and bijective. 
\end{proposition}
Let us define the Hermitian forms $Q_{k}$ on $\coinf(M; V_{k})$: 
\[
 \bar{u_{k}}\dual Q_{k}u_{k}\defeq \i (u_{k}| G_{k}u)_{V_{k}(M)},
\]
and the {\em physical charge}
\[
 Q_{I,2}\defeq Q_{2}\circ I= I^{*}\circ Q_{2}.
 \]
\begin{definition}
The \emph{physical phase space} is  the Hermitian space $(\cV_{P}, Q_{P})$, where:
\[
\cV_{P}= \frac{\Ker_{\c} K^\star}{\Ran_{\c} P}, \quad [\bar{u}]\dual Q_{P}[u]=\bar{u}\dual  Q_{I, 2}u,  \  [u]\in  \frac{\Ker_{\c} K^\star}{\Ran_{\c} P}.
\]
\end{definition}

$Q_{P}$ is a well-defined Hermitian form on $\cV_P$. 

\subsubsection{Phase space of Cauchy data}\label{sec1.6.2}
The following results are proved in \cite{GW2}.
\begin{proposition}\label{prop:rhopasses} The induced map
\[
[\varrho_2]: \ \dfrac{\Ker_{\sc} D_{2}\cap\Ker_{\sc} K^{\star}}{K\Ker_{\sc}D_{1} }\longrightarrow \dfrac{\Ker_{\c} K_{\Sigma}^{\dagger}}{\Ran_{\c} K_\Sigma}
\]
is well defined and bijective.
\end{proposition}

\begin{proposition}\label{prop-added}
 The map 
 \[
 [\varrho_{2}G_{2}]: (\dfrac{\Ker_{\c} K^\star}{\Ran_{\c} P},Q_{P})\longrightarrow(\dfrac{\Ker_{\c} K^{\dagger}_{\Sigma}}{\Ran_{\c} K_\Sigma},q_{I, 2})
 \]
 is an isomorphism of Hermitian spaces.
\end{proposition}
\subsection{Quantization}\label{sec1.7}

 The algebraic quantization of linear gauge theories is discussed in detail in \cite[Sect.~3]{GW2}. The algebraic framework reduces the quantization problem to showing the existence of physically relevant quantum states on the  CCR $*$-algebra $\CCR(\cV_{P}, Q_{P})$ associated to the Hermitian space $(\cV_{P}, Q_{P})$ defined in Subsect. \ref{sec1.6}.   The notions of CCR $*$-algebras, quasi-free states and covariances  are explained for example in \cite[Chap. 4]{G}.

\subsubsection{Covariances}\label{sec1.7.1}
A quasi-free state on $\CCR(\cV_{P}, Q_{P})$ is determined by a pair $\Lambda_{P}^{\pm}$ of {\em covariances}, i.e.~of Hermitian forms on $\cV_{P}$  such that \[
\begin{array}{rl}
i)&\Lambda^{\pm}_{P}= \Lambda_{P}^{\pm*}, \ \Lambda_{P}^{\pm}\geq 0,\\[2mm]
ii)&
\Lambda_{P}^{+}- \Lambda_{P}^{-}= Q_{P}.
\end{array}
\]
We will consider quasi-free states $ \omega$ on  $\CCR(\cV_{P}, Q_{P})$ with covariances obtained  from a  pair of continuous Hermitian forms $\Lambda_{2}^{\pm}$ on $\coinf(M; V_{2})$ (called  the {\em spacetime covariances} of $\omega$) by:
 \beq\label{e00.1}
 [\bar{u}]\dual \Lambda_{P}^{\pm}[u]= \bar{u}\dual \Lambda_{2}^{\pm}u, \ [u]\in \frac{\Ker_{\c} K^\star}{\Ran_{\c} P}.
 \eeq
\begin{lemma}\label{lemma.had}
 Suppose  that $\Lambda_{2}^{\pm}\in L(  \coinf(M; V_{2}), \coinf(M; V_{2})^{*})$ are such that:
  \beq\label{defodefo}
\begin{aligned}
i)&\quad  D_{2}^{*} \circ \Lambda^\pm_{2}=\Lambda^\pm_{2}  \circ D_{2} =0 , \\ 
ii)&\quad \Lambda_2^+-\Lambda^-_2 = Q_{I, 2} \hbox{ on }\Ker K_{\c}^{\star},\\
iii)&\quad \Lambda_{2}^{\pm}=0\hbox{ on }\Ker_{\c} K^{\star}\times \Ran_{\c}K,\\
v) & \quad \Lambda_{2}^{\pm}= \Lambda_{2}^{\pm*}, \ \Lambda^\pm_{2}  \geq 0  \hbox{ on } \Ker_{\c} K^{\star}.
\end{aligned}
\eeq
Then $\Lambda_{2}^{\pm}$ are the covariances  of a quasi-free state on $\CCR(\cV_{P}, Q_{P})$.\end{lemma}
\subsection{Hadamard condition} \label{sec1.7b}
The general consensus is that the  \emph{Hadamard condition}  singles out the physically meaningful states.  We use  the following definition of Hadamard states \cite{SV}. We recall that
 $$
 \cN=\{ (x,\xi)\in T^*M\setminus\zero :  \xi\cdot \rg^{-1}(x)\xi = 0 \}
 $$
 is  the  characteristic set of the wave operator on $(M,\rg)$, and 
$$
\cN^{\pm}=\cN\cap \{(x, \xi)\in T^*M\setminus\zero :\pm v\dual \xi>0 \,\ \forall v\in T_{x}M\hbox{ future-directed time-like}\}
$$
are its  two connected components, corresponding to the upper/lower energy shells.

To formulate the Hadamard condition, we need to identify  the Hermitian forms $\Lambda_{2}^{\pm}$  with distributional kernels  $\lambda_{2}^{\pm}(\cdot, \cdot)\in \cD'(M\times M; L(V_{2}))$, called  {\em two-point functions}.  

This identification is   defined by the formal identity
\[
\bar{u}\dual \Lambda^{\pm}_{2}v\eqdef\int_{M\times M}(u(x)| \lambda_{2}^{\pm}(x, y)v(y))_{V_{2}}\dvol_{\rg}(x)\dvol_{\rg}(y), \ u, v\in \coinf(M; V_{2}).
\]
One can of course use other Hermitian forms on  the fibers of $V_{2}$ to do this identification, like for example $(\cdot| \cdot)_{I, V_{2}}$ or a Hilbertian scalar product as will be done in Sect. \ref{sec.had}. This change amounts  to compose  $\lambda_{2}^{\pm}(x, y)$ by smooth  linear operators acting on the fibers of $V_{2}$ over $x$ and $y$ and does not change the Hadamard condition \eqref{defidefi} below.

\begin{definition}\label{defhadama}
  A quasi-free state $\omega$ on $\CCR(\cV_{P}, Q_{P})$ given by  covariances $\Lambda_{2}^{\pm}$ as in Lemma \ref{lemma.had} is {\em Hadamard} if in addition to \eqref{defodefo} it satisfies:
\beq\label{defidefi}
 \WF(\lambda_{2}^{\pm})'\subset \cN^{\pm}\times \cN^{\pm}.
 \eeq
\end{definition}

\subsubsection{Hadamard condition on a Cauchy surface}\label{sec1.7.3}
One can equivalently consider Hermitian forms  $\lambda^{\pm}_{2\Sigma}$  on  the space of Cauchy data $\coinf(\Sigma; V_{2}\otimes \cc^{2})$ called {\em Cauchy surface covariances}.
Namely   assume that we have a pair of Hermitian forms 
\[
\lambda_{2\Sigma}^{\pm}\in L(\coinf(\Sigma; V_{2}\otimes \cc^{2}), \coinf(\Sigma; V_{2}\otimes \cc^{2})^{*})
\]
 and  set
 \beq\label{e00.2b}
 \Lambda_{2}^{\pm}= (\varrho_{2}G_{2})^{*}\lambda_{2\Sigma}^{\pm}(\varrho_{2}G_{2}).
 \eeq
 The conditions on $\lambda_{2\Sigma}^{\pm}$ corresponding to \eqref{defodefo} are
 \beq
 \label{defodefo-cauchy}
 \begin{aligned}
i)&\quad \lambda_{2\Sigma}^{+}-\lambda_{2\Sigma}^{-} = q_{I, 2} \hbox{ on }\Ker_{\c} K_{\Sigma}^{\dag},\\
ii)&\quad \lambda_{2\Sigma}^{\pm}= 0\hbox{ on }\Ker_{\c} K_{\Sigma}^{\dag}\times \Ran_{\c}K_{\Sigma},\\
iii) & \quad \lambda_{2\Sigma}^{\pm}= \lambda_{2\Sigma}^{\pm*}, \ \lambda^\pm_{2\Sigma}  \geq 0 \mbox{ on } \Ker_{\c} K_{\Sigma}^{\dag}.
\end{aligned}
 \eeq
 
Since $q_{I, 2}$ is non-degenerate, we can set
\beq\label{e00.4}
\lambda^{\pm}_{2\Sigma}\eqdef \pm q_{I, 2}\circ  c_{2}^{\pm}.
\eeq
\begin{proposition}\label{prop:states1} Suppose $c_2^\pm: \coinf(\Sigma;V_{2}\otimes \cc^{2})\to \cinf(\Sigma;V_{2}\otimes \cc^{2})$ is a pair of operators such that:
\beq\label{condit-hada}
\begin{array}{rl}
i)&c_2^++c_2^-=\one,\\[2mm]
ii)&c_2^\pm : \Ran_{\c}K_{\Sigma}\to \Ran K_{\Sigma}, \\[2mm]
iii)&q_{I, 2}\circ  c_{2}^{\pm}= c_{2}^{\pm*}\circ  q_{I, 2}, \ \pm q_{I, 2} \circ  c^\pm_2 \geq 0, \ \hbox{ on } \Ker_{\c} K_{\Sigma}^\dag.\\[2mm]
\end{array}
\eeq

Then $\Lambda_{2}^{\pm}$ given by \eqref{e00.2b} and \eqref{e00.4} are the covariances of a quasi-free state on $\CCR(\cV_{P}, Q_{P})$. Furthermore if for some neighborhood $\cU$ of $\Sigma$ in $M$ we have: 
\[
{\it iv)}\ \WF(U_{2}\circ c_{2}^{\pm})'\subset (\cN^{\pm}\cup\cF)\times T^{*}\Sigma
\] over $\cU\times \Sigma$,
where $\cF\subset T^{*}M$ is a conic set with $\cF\cap \cN= \emptyset$, 
then the associated state  is Hadamard. 
\end{proposition}
The proof of \eqref{condit-hada} is analogous to the one found in \cite[Sect. 3.4]{GW2}. The proof of the statement on the Hadamard condition can be found in \cite[Sect. 11.1]{G}.
\begin{remark}
If $[c_{2}^{\pm}, I_{\Sigma}]=0$ then we can replace  the first condition in \eqref{condit-hada} {\it iii)} by the simpler
\[
q_{2}\circ  c_{2}^{\pm}= c_{2}^{\pm*}\circ  q_{2}.
\]
Conversely 
 if $c_{2}^{\pm}$ satisfy the conditions in Prop. \ref{prop:states1} then  setting
\[
\hat{c}_{2}^{\pm}= \12(c_{2}^{\pm}+ I_{\Sigma}\circ  c_{2}^{\pm}\circ  I_{\Sigma}),
\]
we obtain that $\hat{c}_{2}^{\pm}$ satisfy also the conditions in Prop. \ref{prop:states1} and $[\hat{c}_{2}^{\pm}, I_{\Sigma}]=0$. The only point deserving some attention is the microlocal condition {\it iv)}, which follows from the fact that $\WF(I)'$ and $\WF(I_{\Sigma})'$ are included in the diagonal of $T^{*}M\times T^{*}M$ and $T^{*}\Sigma\times T^{*}\Sigma$ respectively.
\end{remark}

\section{Hadamard covariances}\label{sec.had}
In  \cite[Sect. 5]{GMW} we constructed Hadamard covariances $\lambda_{i\Sigma}^{\pm}$ for the operators $D_{i}$, $i=1,2$ appearing in linearized gravity.  Some properties of these covariances were deduced from Wick rotation, ie $\lambda_{i\Sigma}^{\pm}$ were obtained from \calde projectors associated to elliptic operators $\tD_{i}$ obtained from $D_{i}$ by a Wick rotation in a Gaussian time coordinate associated to some Cauchy surface $\Sigma$.  

This procedure requires analyticity of the metric $\rg$ (or at least partial analyticity in Gaussian time). 
The essential property of the covariances constructed in this way is a positivity property with respect to an Euclidean charge $\tq$ defined in \eqref{e01.3bb}.

In absence of analyticity, one can replace the Wick rotated metric $\trg$ by an almost analytic extension of $\rg$ and obtain the same conclusions.

Another possibility is to prove the positivity property directly, which is what we will do in this section. 
\subsection{The framework}\label{sec.had1}
In order to keep the exposition relatively short, we will adopt the framework   in \cite[Sect. 5]{GMW} to which we refer the reader for notation and proofs. 
\subsubsection{Spacetime and Hermitian bundle}\label{sec.had1.1} 
We set $M= I_{t}\times \Sigma_{\rx}$, where $ I\subset \rr$ is an interval with $0\in \mathring{I}$ and $(\Sigma, \rh_{0})$ a $d$-dimensional Riemannian manifold of bounded geometry. 

Note that in later sections $\Sigma$ will be assumed to be compact, so all the assumptions below related to bounded geometry are automatically satisfied.

We set $\Sigma_{t}= \{t\}\times \Sigma$ and identify $\Sigma_{0}$ with $\Sigma$.
The dual variables to $(t, \rx)$ are denoted by $(\tau, {\rm k})$.

 We fix a $t$-dependent Riemannian metric on $\Sigma$,
 \[
 \rh:I \ni t\mapsto \rh(t)\in \cinfb(I; \BT_{2}^{0}(\Sigma, \rh_{0})).
 \]
  We assume that $\rh(0)= \rh_{0}$ and for ease of notation often denote  $\rh(t)$  by $\rh_{t}$.
 
 We equip $M$ with the Lorentzian metric
 \beq\label{e-03}
 \rg\defeq -dt^{2}+ \rh_{t}d\rx^{2}.
 \eeq
We fix a finite rank complex vector bundle $V\xrightarrow{\pi}\Sigma$ of bounded geometry  over $(\Sigma, \rh_{0})$.  We still denote by $V$  the vector bundle over $M$: $I\times V\xrightarrow{\pi} M$  which is a vector bundle  with the same fibers as $V$.

We denote by $\Diff(M; V)$, resp. $\Diff(\Sigma; V )$ the space of differential operators on $M$ resp. $\Sigma$ acting on sections of $V$.

We assume that  $V\xrightarrow{\pi}M$ is equipped with a non-degenerate fiberwise {\em Hermitian} structure $(\cdot| \cdot)_{V}$, which is  {\em independent} of $t$.
 
 We  fix a reference fiberwise {\em Hilbertian} structure $(\cdot| \cdot)_{\tV}$ on the fibers of $V$ which is also independent of $t$.
 
We write $\tV$ instead of $V$ to emphasize that $V$ is tacitely equipped with the Hilbertian structure $(\cdot| \cdot)_{\tV}$.
 
 We will use this Hilbertian structure to identify sesquilinear forms on the fibers of $V$ with linear operators.

 If $x\in M$ and $u, v\in V_{x}$ we have 
 \[
 (u| v)_{V}= (u| \tau_{x}v)_{\tV},  \ \ \tau_{x}\in L(V_{x}),
 \]
 and we denote by $\tau\in \cinf(M; L(V))$ the corresponding section, which is independent on $t$.
  
 Note that $\tau= \tau^{*}$ and without loss of generality  we can assume that
 \[
 \tau^{*}\tau= \one,\hbox{ i.e. } \tau\hbox{ is unitary for }(\cdot| \cdot)_{\tV}.
 \]
 If $a\in L(V_{x})$ for  $x\in M$ we denote by $a^{*}$, resp.~ $a^{\star}$, the adjoints of $a$ for $(\cdot| \cdot)_{\tV}$, resp.~ $(\cdot| \cdot)_{V}$. Then,
   \begin{equation}
 \label{e01.1}
 a^{\star}= \tau^{-1}_x a^{*}\tau_x \end{equation}
for some $\tau_{x}\in L(V_{x})$.

For $u, v\in \cinf_{\rm sc}(M; V)$ we set
 \[
 \begin{array}{l}
  (u|v)_{\tV(\Sigma_{t})}\defeq\int_{\Sigma_{t}}(u|v)_{\tV}|\rh_{0}|^{\12}d\rx,\\[2mm]
  (u|v)_{V(\Sigma_{t})}\defeq\int_{\Sigma_{t}}(u|v)_{V}|\rh_{0}|^{\12}d\rx= (u| \tau v)_{\tV(\Sigma_{t})},\\[2mm]
  (u| v)_{\tV(M)}\defeq\int_{M}(u(t)|v(t))_{\tV}|\rh_{0}|^{\12}dtd\rx,\\[2mm]
 (u| v)_{V(M)}\defeq\int_{M}(u(t)|v(t))_{V}|\rh_{0}|^{\12}dtd\rx= (u| \tau v)_{\tV(M)}.
\end{array}
 \]
If $\Omega\subset M$ is some open set, we  also denote 
\[
\begin{array}{l}
 (u| v)_{\tV(\Omega)}\defeq\int_{\Omega}(u(t)|v(t))_{\tV}|\rh_{0}|^{\12}dtd\rx,\\[2mm]
 (u| v)_{V(\Omega)}\defeq\int_{\Omega}(u(t)|v(t))_{V}|\rh_{0}|^{\12}dtd\rx= (u| \tau v)_{\tV(\Omega)}.
\end{array}
\]
We denote by $L^{2}(\Sigma; \tV)$ the $L^{2}$ space obtained from the Hilbertian scalar product $(\cdot| \cdot)_{\tV(\Sigma)}$.
 \subsubsection{Adjoints}\label{sec01.2.2}
 If  $a\in \cinfb(I; \Diff(\Sigma; V))$, resp.~ $A\in \Diff(M; V)$, we denote by  $a^{*}$ resp.~ $A^{*}$ its formal adjoint for $(\cdot| \cdot)_{\tV(\Sigma_{t})}$ resp.~$(\cdot| \cdot)_{\tV(M)}$.  We set $\Re a= \12 (a+ a^{*})$. 
 
 We denote by $a^{\star}$ resp.~ $A^{\star}$ its formal adjoint for $(\cdot| \cdot)_{V(\Sigma_{t})}$ resp.~$(\cdot| \cdot)_{V(M)}$. As above we have:
 \[
  a^{\star}= \tau^{-1} a^{*}\tau, \quad A^{\star}= \tau^{-1}A^{*}\tau.
 \]

 \subsubsection{Hyperbolic operator}\label{sec01.2.3}
 We fix a $t$-dependent differential operator $a= a(t, \rx, D_{\rx})$ belonging to $\cinfb(I; \Diff^{2}(\Sigma; V))$ and  denote by $\sigma_{\rm pr}(a)\in \cinf(T^{*}\Sigma; L(V))$ its principal symbol.
 
 We assume the following properties:
\[
 \begin{array}{rl}
{\rm (H1)}& \ a(t) = a^{\star}(t),\  t\in I, \\[2mm]
{\rm (H2)}&\sigma_{\rm pr}(a)(t)(\rx, {\rm k})= {\rm k}\dual \rh_{t}^{-1}(\rx){\rm k}\,\one_{V}, \ t\in I.
\end{array}
\]
We set
\beq\label{e01.1b}
D\defeq \p_{t}^{2}+ a(t)\hbox{ acting on }\coinf(M; V),
\eeq
which is a hyperbolic operator with scalar principal part. Note that 
 $D= D^{\star}$, but of course $D\neq D^{*}$ in general.

The Cauchy problem
 \begin{equation}
 \label{e01.3b}
\begin{cases}
 Du= 0\hbox{ in }M\\
 \varrho u= f\in  \coinf(\Sigma; V\otimes \cc^{2})
\end{cases}
 \end{equation}
 is well-posed, where  
 \[
 \varrho u= \col{u(0)}{\i^{-1}\p_{t}u(0)}.
 \]
 We denote by $u= Uf$ the unique solution of \eqref{e01.3b}. We set:
  \begin{equation}
 \label{e01.3bb}
 \tq\defeq \mat{0}{1}{1}{0}, \ q\defeq \mat{0}{\tau}{\tau}{0}.
 \end{equation}
 The Cauchy evolution for $D$ is pseudo-unitary for $q$.
 \subsection{Hadamard projectors}\label{sec.had2}
In \cite[Sect. 5]{GMW} we constructed projectors $c^{\pm}$, acting on $\coinf(\Sigma; \tV\otimes \cc^{2})$, called {\em Hadamard projectors} such that
 \beq\label{e-01}
 \begin{array}{rl}
1)&c^{+}+ c^{-}= \one,\\[2mm]
2)&qc^{\pm}= c^{\pm*}q,\\[2mm]
3)&\WF(U\circ c^{\pm})'\subset (\cN^{\pm}\cup \cF)\times T^{*}\Sigma\hbox{  for }\cF= \{k=0\}\subset T^{*}M.
 \end{array}
 \eeq
We define Cauchy surface covariances by \[
\bar{f}\dual \lambda_{\Sigma}^{\pm}f= \pm (f|q c^{\pm}f)_{\tV(\Sigma)\otimes \cc^{2}}, 
\]
and the associated spacetime covariances by
\[
\Lambda^{\pm}= (\varrho \circ G)^{*}\lambda_{\Sigma}^{\pm}(\varrho\circ G),
\]
where $G$ is the causal propagator for $D$. The charge $Q$ is
\[
\bar{u}\dual Q u= \i (u| Gu)_{V(M)}, \ u\in \coinf(M; V).
\]
We have then
\[
\begin{array}{rl}
i)& D^{*}\Lambda^{\pm}= \Lambda^{\pm}D=0, \Lambda^{+}- \Lambda^{-}= Q,\\[2mm]
ii)&\Lambda^{\pm}= \Lambda^{\pm*},\\[2mm]
iii)&\WF(\lambda^{\pm})'\subset \cN^{\pm}\times \cN^{\pm}.
\end{array}
\]
In other words, except for the positivity, $\Lambda^{\pm}$, $\lambda^{\pm}_{\Sigma}$ are Hadamard covariances for $D$.

The following lemma states the positivity of $c^{\pm}$ for the Euclidean charge $\tq$ defined in 
 \eqref{e01.3bb}.
 \begin{lemma}\label{lemma.e-01}
 We can construct the Hadamard projectors $c^{\pm}$ so that in addition to \eqref{e-01} one has
\begin{equation}
\label{e-02}
\pm(c^{\pm} f| \tq c^{\pm}f)_{\tV(\Sigma)\otimes \cc^{2}}\geq 0, \ \forall f\in \coinf(\Sigma;\tV\otimes \cc^{2}).
\end{equation}
Moreover  \beq\label{e-05}
\pi_{1}c^{\pm}f=b^{\pm}\pi_{0}c^{\pm}f\hbox{ for }\pi_{i}f= f_{i}, \ f= \col{f_{0}}{f_{1}}, 
\eeq
 where $b^{+}= b$, $b^{-}= - b^{\star}$ and $b$ is an elliptic  first order pseudodifferential operator such that $b: H^{s}(\Sigma; \tV)\to H^{s-1}(\Sigma; \tV)$ is an isomorphism for any $s\in \rr$.
  \end{lemma}
\proof We use the notation in \cite[5.2.7-5.2.10]{GMW}. The projectors $c^{\pm}$ are given by
\beq\label{e-12}
c^{\pm}= T \pi^{\pm}T^{-1}, \ \pi^{+}= \mat{1}{0}{0}{0}, \ \pi^{-}= \mat{0}{0}{0}{1},
\eeq
where $T$ is defined in \cite[equ. (5.13)]{GMW}. 

A concrete expression for $c^{\pm}$ is:
\beq\label{e-04}
c^{\pm}=  \mat{\mp(b^{+}- b^{-})^{-1}b^{\mp}}{\pm(b^{+}- b^{-})^{-1}}{\mp b^{\pm}(b^{+}- b^{-})^{-1}b^{\mp}}{\pm b^{\pm}(b^{+}- b^{-})^{-1}},
\eeq
where  $b^{+}= b$, $b^{-}= - b^{\star}$ and $b$ is an elliptic first order pseudodifferential operator on $\Sigma$.

To prove that $\pm c^{\pm*}\tq c^{\pm}\geq 0$, we need to compute  $T^{*}\tq T$. For the operator $b$ constructed in \cite[Prop. 5.2]{GMW} and $c$ defined in \cite[5.2.9]{GMW} we have
\[
T= S\circ\mat{c}{0}{0}{c}, \ S= -\i \mat{1}{-1}{b}{\tau^{*}b^{*}\tau}(b+ \tau^{*}b^{*}\tau)^{-1}.
\]
We compute $T^{*}$ and using that $\tq= \mat{0}{1}{1}{0}$, we obtain that
\[
T^{*}\tq T= (c(b+ \tau^{*}b^{*} \tau)^{-1})^{*}\mat{b+ b^{*}}{- b^{*}+ \tau^{*}b^{*}\tau}{-b+ \tau^{*}b\tau}{- \tau^{*}(b+ b^{*})\tau}(c(b+ \tau^{*}b^{*} \tau)^{-1}).
\]
Therefore using \eqref{e-12} we see that $\pm c^{\pm*}\tq c^{\pm}\geq 0$ iff $b+ b^{*}\geq 0$. The construction of $b$ is given in \cite[Prop. 5.2]{GMW}. Concretely we have:
\[
b= \epsilon+ b_{0}(1- \chi_{R}(a_{\rm ref})),
\]
$R\gg 1$ being a large parameter. The first order elliptic pseudodifferential operator $\epsilon$ satisfies $\Re \epsilon\geq 1$. Therefore 
\[
b+b^{*}= (2 \Re  \epsilon)^{\12}(1- \tilde{s}_{-1})(2\Re \epsilon)^{\12},
\]
where 
\[
\tilde{s}_{-1}= -(2 \Re \epsilon)^{-\12}(b_{0}(1- \chi_{R}(a_{\rm ref})+(1- \chi_{R}(a_{\rm ref}))b_{0}^{*})(2\Re \epsilon)^{-\12}.
\]
As in the proof of \cite[Prop. 5.2]{GMW} the norm of $\tilde{s}_{-1}$ in $B(L^{2}(\Sigma; \tV))$ tends to $0$ when $R\to +\infty$, so choosing $R\gg 1$ we obtain that $b+b^{*}\geq c>0$. This completes the proof of the first statement of the lemma.

The second statement follows from \eqref{e-04}, except for the fact that $b$ can be chosen invertible. The principal symbol of    $b$ is $(k\dual \rh_{0}(\rx)k)^{\12}\one$, so  $b$ is elliptic.  We have seen above that we can choose $R\gg 1$  so that $b+ b^{*}\geq c>0$. The sesquilinear form associated to $b$ with domain $H^{\12}(\Sigma; \tV)$ is closed and coercive, so $b= H^{\12}(\Sigma; \tV)\to H^{-\12}(\Sigma; \tV)$ is an isomorphism. This extends to any $s\in \rr$ by the usual argument.   \qed
\subsection{Application to $D_{k}$}\label{sec.had.1}
Let us now  recall how to apply the previous constructions to the operators $D_{k}$, $k= 1, 2$.

If $(M, \rg)$ is globally hyperbolic and $\Sigma$ is a smooth spacelike Cauchy surface, then if $(M, \rg)$ is of {\em bounded geometry} near $\Sigma$, see \cite[Def. 3.2]{GMW} for the precise definition, then using Gaussian normal coordinates to $\Sigma$  one can isometrically map a neighborhood of $\Sigma$ in $M$ to $I\times \Sigma$, equipped with  a metric as in \eqref{e-03}.   The bounded geometry assumption is automatically satisfied if $\Sigma$ is compact, as we will assume in later sections.

\subsubsection{Reduced setting}\label{sec.had.1.1}
Conjugating $D_{k}$ by   an isomorphism corresponding to parallel transport along $\p_{t}$ one can then reduces oneself to the situation in  Subsect. \ref{sec.had1}, see \cite[Subsect. 4.4]{GMW}. For $k=2$ this isomorphism maps the background metric $\rg$ to $\rg_{0}= - dt^{2}+ \rh_{0}$.

In this reduced setting, several operators take simpler forms: for example we have
\[
\begin{array}{l}
D_{k}= \p_{t}^{2}+ a_{k}(t, \rx, \p_{\rx}),\\[2mm]
Iu_{2}= u_{2} - \frac{1}{4}(\rg_{0}| u_{2})_{V_{2}},
\end{array}
\]
and the expression of $d$ can be found in \cite[Prop. 4.11]{GMW}. The operators $\tau_{k}$ relating the Hermitian and Hilbertian structures on $V_{k}$ are
\[
\tau_{1}= \mat{-1}{0}{0}{-1}, \ \tau_{2}= \left(\begin{array}{ccc}
1&0&0\\0&-1&0\\
0&0&1
\end{array}\right),
\]
where we use the decompositions of $(0, k)$-tensors recalled in \ref{sec1.1.3}.

\subsubsection{Gauge invariance modulo smooth errors}\label{sec.had.1.2}
To complete this subsection, we write an easy lemma. We state and prove it only in the case when $\Sigma$ is compact, but  the result extends easily  to the bounded geometry framework.

\begin{lemma}\label{blip}
Assume that $\Sigma$ is compact. Let $c_{k}^{\pm}$ for $k=1,2$ be Hadamard projectors for $D_{k}$. Then 
\[
c_{2}^{\pm}K_{\Sigma}= K_{\Sigma}c_{1}^{\pm} \pm r_{-\infty},
\]
where $r_{-\infty}= c_{2}^{+}K_{\Sigma}c_{1}^{-}- c_{2}^{-}K_{\Sigma}c_{1}^{+}\in \Psi^{-\infty}(\Sigma)$ is smoothing.
\end{lemma}
\proof Let $f_{1}\in \cE'(\Sigma; V_{1}\otimes \cc^{2})$,  $u_{1}= U_{1}c_{1}^{+}f_{1}$, $u_{2}= K u_{1}$ and $f_{2}=\varrho_{2}u_{2}= K_{\Sigma}c_{1}^{+}f_{1}$. We have $\WF u_{1}\subset \cN^{+}$  hence $\WF u_{2}\subset \cN^{+}$ since $K$ is a differential operator. On the other hand $\WF U_{2}c_{2}^{\pm}f_{2}\subset \cN^{\pm}$ hence $\WF U_{2}c_{2}^{-}f_{2}\subset \cN^{+}\cap \cN^{-}= \emptyset$ since $c_{2}^{-}= 1- c_{2}^{+}$. Therefore $c_{2}^{-}K_{\Sigma}c_{1}^{+}$ is smoothing \qed

\subsubsection{Additional symmetry}\label{sec.had.1.3}The operator $D_{2}$ has the additional symmetry $[I, D_{2}]=0$, which can be carried over to $b_{2}$ or the Hadamard projectors $c_{2}^{\pm}$. 

In fact if $b_{2}$ is the operator entering in the construction of $c_{2}^{\pm}$, we see that $I\circ b_{2}\circ I$ also satisfies the  conditions in \cite[Prop. 5.2]{GMW}, which caracterize $b$ uniquely modulo a smoothing error. Therefore we can replace $b_{2}$ by $\12 (b_{2}+ I\circ b_{2}\circ I)$ modulo a smoothing error and assume that $[I, b_{2}]=0$.  

This replacement does not invalidate the properties of $c_{2}^{\pm}$ summarized in Subsect. \ref{sec.had2}.  The projectors $c_{2}^{\pm}$ have now the additional property
\[
[I_{\Sigma}, c_{2}^{\pm}]=0.
\]
Therefore $c_{2}^{\pm}$ are also selfadjoint for the physical Hermitian form $q_{I, 2}= q_{2}\circ I_{\Sigma}$ ie
\beq\label{e-07}
(f_{2}| q_{2}I_{\Sigma}c_{2}^{\pm}f_{2})_{\tV_{2}\otimes \cc^{2}}= (c_{2}^{\pm}f_{2}| q_{2}I_{\Sigma}f_{2})_{\tV_{2}\otimes \cc^{2}}.
\eeq
 
 \section{Gauge fixing}\init\label{sec10}
\def\tV{\tilde{V}}\subsection{Introduction}\label{sec10.1}
To motivate the constructions in this section we start with   some comments on conditions \eqref{defodefo} and  \eqref{defidefi} in Subsect. \ref{sec1.7} on covariances  generating a quasi-free state on $\CCR(\cV_{P}, Q_{P})$.

For a quantum field theory associated to $D_{2}$, one usually assumes condition \eqref{defodefo} {\it i)} and $\lambda_{2}^{+}- \lambda_{2}^{-}= \i G_{2}$, ie \eqref{defodefo} {\it ii)} extended to the larger space $\coinf(M; V_{2})$. Together with \eqref{defidefi}, these two conditions  fix $\lambda_{2}^{\pm}$ uniquely, modulo smooth kernels. 

The positivity condition \eqref{defodefo} {\it v)} is in general  not satisfied on arbitrary test fields, but only on $\Ker_{\c}K^{\star}$. This comes from the fact that the fiber scalar product $(\cdot| \cdot)_{V_{2}}$ for which $D_{2}$ is formally selfadjoint is non positive.

The condition \eqref{defodefo} {\it iv)} that $\lambda_{2}^{\pm}$ map $\Ran_{\c}K$ into $\Ran K$  is the crucial {\em gauge invariance condition}, which  implies that $\Lambda^{\pm}$ are well defined on the physical phase space $\frac{\Ker_{\c} K^\star}{\Ran_{\c} P}$. This condition is the most difficult to impose.

A way out of this difficulty is to try to eliminate  the remaining gauge freedom.  

Working for example with the phase space $\dfrac{\Ker_{\sc} D_2\cap\Ker_{\sc} K^\star}{K\Ker_{\sc}D_{1}}$ this amounts to impose more  gauge fixing conditions in addition to the harmonic gauge condition $K^{\star}u_{2}=0$ in order to eliminate the remaining gauge freedom corresponding to $K\Ker_{\sc}D_{1}$.

The additional gauge fixing conditions should hence uniquely specify a supplementary space to $K\Ker_{\sc}D_{1}$ inside  $\Ker_{\sc} D_2\cap\Ker_{\sc} K^\star$. These additional gauge fixing conditions  should moreover be chosen so that the positivity condition on the covariances is satisfied on this supplementary space.

\subsubsection{TT gauge condition}\label{sec10.1.1}
A first way to reduce the gauge freedom is to impose the {\em traceless condition} by requiring that $(\rg| u_{2})_{V_{2}}= 0$. Together with  in  $K^{\star}u_{2}=0$, this  is called the {\em transverse-traceless} (TT){ \em gauge}.

This is always possible if $\Lambda\neq 0$ see eg \cite[Thm. 2.7]{FH}.  Setting 
\[
K_{0}: \cinf(M; V_{0})\ni u_{0}\mapsto u_{0}\rg\in \cinf(M; V_{2}),
\]
we have $K_{0}^{\star}u_{2}= - (\rg| u_{2})_{V_{2}}$ and one can show  that
 \beq\label{e10.-2}
[Id]: \frac{\Ker_{\sc} D_{2}\cap \Ker_{\sc} K^{\star}_{}\cap \Ker_{\sc} K_{0}^{\star}}{K\Ker_{\sc} D_{1}\cap \Ker_{\sc} K_{0}^{\star}}\longrightarrow \frac{\Ker_{\sc} P}{\Ran_{\sc} K}
\eeq
is again an isomorphism.  The key fact is that 
\begin{equation}
\label{e10.-1}
K_{0}D_{0}= D_{2}K_{0},
\end{equation}
  by Prop. \ref{prop0.-1}, where $D_{0}= - \Box_{0}-2\Lambda$ was introduced in Subsect. \ref{sec1.5}.

This does not fully eliminate the gauge freedom, ie we still have a quotient space in \eqref{e10.-2}.  
\subsubsection{Synchronous gauge condition}\label{sec10.1.2}
A  possibility is to impose the {\em synchronous gauge} condition.  After fixing a Cauchy surface $\Sigma$ and introducing Gaussian normal coordinates $(t, \rx)$ to  $\Sigma$, one requires that $u_{t\Sigma}=0$ near $\Sigma$. It is shown in \cite[Thm. 2.8]{FH} that  for any $u_{2}\in \cinf(M; V_{2})$ such that $Pu_{2}=0$, there exists $u_{1}\in \cinf(M; V_{1})$, such that $(u_{2}- K u_{1})_{t\Sigma}=0$ near $\Sigma$.  However $K^{\star}Ku_{1}$ does not necessarily vanish, ie the harmonic gauge condition $K^{\star}u_{2}=0$ is destroyed by this gauge transformation.

One can weaken the synchronous gauge condition by requiring only that
\beq\label{e10.1}
u_{2t\Sigma}\traa{\Sigma}=0, \nabla_{\nu}u_{2t\Sigma}\traa{\Sigma}=0.
\eeq
Note that unlike the harmonic and traceless gauge conditions,  the above condition does not 'propagate' to the whole spacetime $M$, because  in general $u_{2t\Sigma}$ does not solve a hyperbolic equation, even if $D_{2}u_{2}$=0.

One can ask if is possible to impose \eqref{e10.1}, together with the TT gauge condition $K^{\star}u_{2}= (g|u_{2})_{V_{2}}=0$.  One can call this the {\em TT-synchronous} gauge condition. 

Given $u_{2}\in \Ker_{\sc} D_{2}\cap \Ker_{\sc} K^{\star}$, we need to find $u_{1}\in \cinf(M; V_{1})$ such that
\beq\label{e10.2}
\left\{
\begin{array}{rl}
 &D_{1}u_{1}=0, \\[2mm]
1)&(\rg| K u_{1})_{V_{2}}\traa{\Sigma}= (\rg|u_{2})_{V_{2}}\traa{\Sigma},\\[2mm]
2)&(\rg| \nabla_{\nu}K u_{1})_{V_{2}}\traa{\Sigma}= (\rg|\nabla_{\nu}u_{2})_{V_{2}}\traa{\Sigma},\\[2mm]
3)&(Ku_{1})_{t\Sigma}\traa{\Sigma}= u_{2t\Sigma}\traa{\Sigma},\\[2mm]
4)&(\nabla_{\nu}Ku_{1})_{t\Sigma}\traa{\Sigma}= \nabla_{\nu}u_{2t\Sigma}\traa{\Sigma}.
\end{array}
\right.
\eeq
If $\tilde{u}_{2}= u_{2}- Ku_{1}$, then $K^{\star}u_{2}=0$, the initial conditions (1)  and (2) ensure that $(\rg| \tilde{u}_{2})_{V_{2}}=0$, using \eqref{e10.-1}, while (3) and (4) ensure that $\tilde{u}_{2}$ satisfies \eqref{e10.1}. 

 The system \eqref{e10.2} can be rewritten as an elliptic system of equations in terms of the Cauchy data $f_{1}= \varrho_{1}u_{1}$. 
 
If $\Sigma$ is  compact, this elliptic system  is Fredholm.  Except for a finite dimensional subspace in $\Ker D_{2 }$ \eqref{e10.2} has a unique solution,  modulo a finite dimensional subspace in $\Ker D_{1} $.

 \subsection{Microlocal TT-synchronous gauge}\label{sec10.2}
 Even if we ignore the problem with  the possible non invertibility of \eqref{e10.2}, the TT-synchronous gauge condition is not convenient for the construction of gauge invariant Hadamard states for linearized gravity.  In particular states constructed using the TT-synchronous gauge fixing will in general not be Hadamard states.

 It turns out that it is much better to adapt it to the Hadamard  projectors $c_{2}^{\pm}$ for $D_{2}$. Let us now define this modified gauge condition.

  In the rest of the section we assume that  the Cauchy surface $\Sigma$ is compact.

  As before we use Gaussian normal coordinates to $\Sigma$ to isometrically identify a neighborhood of $\Sigma$ in $M$ with $I_{t}\times \Sigma_{\rx}$ for some interval $I\ni 0$, equipped  the metric $- dt^{2}+ h_{t}(\rx)d\rx^{2}$. Under this identification $\Sigma$ is identified with $\{t=0\}$.
 
 We denote by $l: \cinf(\Sigma; V_{2})\to \cinf(\Sigma; V_{1})$ the map:
 \[
 lu_{2}\defeq \col{\frac{1}{2}(\rg_{0}| u)_{V_{2}}\traa{\Sigma}}{ 2 (u_{2t\Sigma})\traa{\Sigma}},
 \]
 where in the rhs we identify as usual  $(v_{t}, v_{\Sigma})\in \cinf(\Sigma; \cc\oplus T^{*}\Sigma)$ with $v_{t}dt+ v_{\Sigma}\in \cinf(\Sigma; V_{1})$.
 
 If $c_{2}^{\pm}$ are the Hadamard projectors whose construction is recalled in Sect. \ref{sec.had} we set for $f_{2}\in \cinf(\Sigma; V_{2}\otimes \cc^{2})$
 \begin{equation}
 \label{e10.3}
 R_{\Sigma} f_{2}\defeq \col{l \pi_{0}c_{2}^{+}f_{2}}{l\pi_{0}c_{2}^{-}f_{2}}\in \cinf(\Sigma; V_{1}\otimes \cc^{2}).
 \end{equation}
 Here $\pi_{0}: V_{2}\otimes \cc^{2}\to V_{2}$ denotes the projection on the first component. 
\begin{definition}\label{def10.1}
 The {\em microlocal TT-synchronous gauge} condition is defined (in terms of Cauchy data on $\Sigma$) by:
 \beq\label{e10.-3}
 \begin{array}{rl}
 (1)&K_{\Sigma}^{\dag}f_{2}=0,\\[2mm]
 (2)&R_{\Sigma}f_{2}=0.
 \end{array}
 \eeq
 \end{definition}

\subsection{Properties of $R_{\Sigma}$ and $R_{\Sigma}K_{\Sigma}$}\label{sec10.3}
Imposing  \eqref{e10.-3} by a gauge transformation  is equivalent to find $f_{1}\in \cinf(\Sigma; \tilde{V}_{1}\otimes \cc^{2})$ solving
\beq\label{e10.-4b}
R_{\Sigma}K_{\Sigma}f_{1}= R_{\Sigma}f_{2},
\eeq
for some given $f_{2}\in \cinf(\Sigma; \tilde{V}_{2}\otimes \cc^{2})\cap \Ker K_{\Sigma}^{\dag}$, so that $f_{2}- K_{\Sigma}f_{1}$ satisfies \eqref{e10.-3}. In this subsection we study the operator $R_{\Sigma}K_{\Sigma}$ appearing in \eqref{e10.-4b} and we prove an important positivity property of the microlocal TT-synchronous gauge condition.

\subsubsection{Some equivalent norms}\label{sec10.3.1}


The convenient Sobolev spaces for Cauchy data   are:
 \[
 \cH^{s}(\Sigma; \tV_{k}\otimes \cc^{2})= H^{s}(\Sigma; \tV_{k})\oplus H^{s-1}(\Sigma; \tV_{k}), \ s\in \rr
 \] 
 equipped with the norm
\[
\|f\|_{s}^{2}= \|f_{0}\|^{2}_{H^{s}(\Sigma; \tV_{k})}+ \|f_{1}\|^{2}_{H^{s-1}(\Sigma; \tV_{k})}, \ f= \col{f_{0}}{f_{1}}.
\]
\begin{lemma}
 The map:
\beq\label{e10.-5}
L_{k}:\begin{array}{l}
\cH^{s}(\Sigma; \tV_{k}\otimes \cc^{2})\to H^{s}(\Sigma; \tV_{k}\otimes \cc^{2})\\[2mm]
 f\mapsto L_{k}f= \col{\pi_{0}c_{k}^{+}f}{\pi_{0}c_{k}^{-}f}\eqdef\col{v^{+}}{v^{-}}=v
\end{array}
\eeq
is an isomorphism.
\end{lemma}
We will  denote by $\pi^{\pm}: \cinf(\Sigma; \tV_{k}\otimes \cc^{2})\to \cinf(\Sigma; \tV_{k})$ the maps $v= \col{v^{+}}{v^{-}}\mapsto v^{\pm}$. 

It is often more convenient to use $\col{v^{+}}{v^{-}}$ instead of $f$.


\proof 
From the expression \eqref{e-04}  of $c^{\pm}$, we deduce that $c^{\pm}$ are bounded on $\cH^{s}$ and  since $c^{+}+ c^{-}= \one$, an equivalent norm on $\cH^{s}$ is given by
\[
\| f\|^{2}_{s, {\rm mod}}= \| c^{+}f\|^{2}_{s}+ \| c^{-}f\|^{2}_{s}.
\]
Using the second statement in Lemma \ref{lemma.e-01}, we obtain that  an equivalent norm is 
\[
(\|v^{+}\|^{2}_{H^{s}(\Sigma; \tV_{i})}+ \| v^{-}\|^{2}_{H^{s}(\Sigma; \tV_{i})})^{\12}, \ v^{\pm}= \pi_{0}c^{\pm}f,
\]
which proves the lemma. \qed

\subsubsection{Properties of $R_{\Sigma}K_{\Sigma}$}\label{sec10.3.2}
\begin{proposition}\label{prop10.1}
\ben
\item 
$R_{\Sigma}K_{\Sigma}: \cH^{s}(\Sigma; \tilde{V}_{1}\otimes \cc^{2})\to  H^{s-1}(\Sigma; \tilde{V}_{1}\otimes \cc^{2})$
is Fredholm of index $0$ for any $s\in \rr$;
\item  $\pi^{\pm}R_{\Sigma}c_{2}^{\mp}=0$;
\item $\pi^{\pm}R_{\Sigma}K_{\Sigma}c_{2}^{\mp}$ is smoothing.
\een
\end{proposition}

\proof 
From  Lemma \ref{blip} we obtain that
\beq\label{e10.-55}
R_{\Sigma}K_{\Sigma}= \mat{l \pi_{0}K_{\Sigma} c_{1}^{+}}{0}{0}{l \pi_{0}K_{\Sigma} c_{1}^{-}}+ R_{-\infty},
\eeq
where $R_{-\infty}$ is a smoothing operator.

We recall that $K_{\Sigma}$ decomposes as $K_{\Sigma}= I_{\Sigma}\circ T_{\Sigma}$, corresponding to $K= I\circ d$. Since $I\rg_{0}= - \rg_{0}$ and $I= I^{\star}$, we obtain that
\[
l\pi_{0}K_{\Sigma}f_{1}= \col{- \frac{1}{4} (\rg_{0}| \pi_{0}T_{\Sigma}f_{1})_{V_{2}}}{ 2 (\pi_{0}T_{\Sigma}f_{1})_{t\Sigma}}.
\]
We recall that we work in the reduced setting, explained in \ref{sec.had.1.1}.

If $w$ is the solution of $D_{1}w=0$ with $\rho w= f_{1}$, then  from \cite[Prop. 4.11]{GMW}  we have:
\beq\label{e10.5}
 \begin{array}{rl}
 - \frac{1}{4} (\rg_{0}| \pi_{0}T_{\Sigma}f_{1})_{V_{2}}=& (dw)_{tt}-\frac{1}{2}(\rh_{0}| (dw)_{\Sigma\Sigma})_{V_{2\Sigma}}\\[2mm]
 =&\p_{t}w_{t}+ \delta_{\Sigma}w_{\Sigma}+ \12 {\rm tr}({\bf r}_{0})w_{t},\\[2mm]
 2(\pi_{0}T_{\Sigma}f_{1})_{t\Sigma}=& 2(dw)_{t\Sigma}= \p_{t}w_{\Sigma}- \12 {\rm tr}({\bf r}_{0})w_{\Sigma}- {\bf r}_{0}w_{\Sigma}+ d_{\Sigma}w_{t},
 \end{array}
\eeq
where ${\bf r}_{0}= \12 \p_{t}\rh_{0}\rh_{0}^{-1}$ and $d_{\Sigma}, \delta_{\Sigma}$ are the symmetric differential and co-differential on $(\Sigma, \rh_{0})$.

By Lemma \ref{lemma.e-01}, we know that  if $f_{1}= c_{1}^{\pm}f_{1}$, then $\p_{t}w\traa{\Sigma}= \i b^{\pm}w\traa{\Sigma}$. 
Therefore 
\beq\label{e10.6}
l\pi_{0}K_{\Sigma}c_{1}^{\pm}f_{1}= (\i b^{\pm}+ B)\pi_{0}c_{1}^{\pm}f_{1}
\eeq
where 
\[
Bv= \mat{\12 {\rm tr}({\bf r}_{0})}{\delta_{\Sigma}}{d_{\Sigma}}{- \12 {\rm tr}({\bf r}_{0})- \12 {\bf r}_{0}}v, \ v= \col{v_{t}}{v_{\Sigma}}\in H^{s}(\Sigma; \tilde{V}_{1}).
\]
Summarizing we obtain
\beq\label{e10.-6}
R_{\Sigma}K_{\Sigma}= \mat{(\i b^{+}+B)\pi_{0} c_{1}^{+}}{0}{0}{(\i b^{-}+B)\pi_{0} c_{1}^{-}}+ R_{-\infty},
\eeq

By Lemma \ref{lemma.e-01}, $b^{\pm}$ are elliptic and invertible and  $\pm \Re b^{\pm}\geq C (-\Delta_{\tilde{h}_{0}}+ 1)^{\12}$. The operator $B$ belongs to $\Psi^{1}(\Sigma; \tilde{V}_{1})$ and is formally selfadjoint. Therefore  the maps $(\i b^{\pm}+ B): H^{s}(\Sigma; \tilde{V}_{1})\to H^{s-1}(\Sigma; \tilde{V}_{1})$ are boundedly invertible,  which implies  (1) since $R_{-\infty}$ is smoothing.  (2) is obvious and (3) follows from \eqref{e10.-6}. \qed

\subsubsection{Positivity property}
\begin{lemma}\label{lemma.e-02}
 We have
 \[
 \pm q_{I, 2}\circ c_{2}^{\pm}\geq 0 \hbox{ on }\Ker R_{\Sigma}.
 \]
\end{lemma}
\proof
Using \eqref{e-07} and the fact that $c_{2}^{\pm}$ are projections  we obtain that
\[
\pm (f_{2}| q_{I, 2}c_{2}^{\pm}f_{2})_{\tV_{2}\otimes \cc^{2}}= \pm (c_{2}^{\pm}f_{2}| q_{I, 2}c_{2}^{\pm}f_{2})_{\tV_{2}\otimes \cc^{2}},
\]
and if $R_{\Sigma}f_{2}=0$ we obtain that 
\[
 \pm (c_{2}^{\pm}f_{2}| q_{I, 2}c_{2}^{\pm}f_{2})_{\tV_{2}\otimes \cc^{2}}=  \pm (c_{2}^{\pm}f_{2}| \tilde{q}_{2}c_{2}^{\pm}f_{2})_{\tV_{2}\otimes \cc^{2}}.
\]
This is positive by Lemma \ref{lemma.e-01}. \qed

\subsection{Gauge fixing in the regular case}\label{sec10.4}
In order to lighten notation, we   often write in this subsection
$\bar{f}\dual q f$ for $(f| q f)_{\tilde{V}\otimes \cc^{2}}$, $\cH^{s}$ for  $\cH^{s}(\Sigma; \tilde{V}\otimes \cc^{2})$, $H^{s}$ for $H^{s}(\Sigma; \tilde{V}\otimes \cc^{2})$.

We assume in this subsection that  $R_{\Sigma}K_{\Sigma}: \cH^{s}(\Sigma; \tilde{V}_{1}\otimes \cc^{2})\to H^{s-1}(\Sigma; \tilde{V}_{1}\otimes \cc^{2})$ is {\em invertible}.  We set
\begin{equation}
\label{e10.7}
T= 1- K_{\Sigma}(R_{\Sigma}K_{\Sigma})^{-1}R_{\Sigma}.
\end{equation}
\begin{proposition}\label{prop10.2}
 \ben
 \item $T: \cH^{s}(\Sigma; \tilde{V}_{2}\otimes \cc^{2})\to \Ker R_{\Sigma}$  is a bounded projection  on $\cH^{s}(\Sigma; \tilde{V}_{2}\otimes \cc^{2})$,
 \item  $T\circ K_{\Sigma}=0$,
\item $T$ preserves $\Ker K_{\Sigma}^{\dag}$.
\item $c_{2}^{\pm}T c_{2}^{\mp}$ is smoothing.
\een
\end{proposition}
\proof
From \eqref{e-04} we obtain that $R_{\Sigma}: \cH^{s}(\Sigma; V_{2}\otimes \cc^{2})\to H^{s}(\Sigma; V_{1}\otimes \cc^{2})$ is bounded. Using the 
 ellipticity of $R_{\Sigma}K_{\Sigma}$ we obtain that $T$ is bounded. We have $R_{\Sigma}\circ T=0$  and $T= \one$ on $\Ker R_{\Sigma}$ which implies (1).  (2)  and   (3) are clear.

To prove (4) is suffices to show that $c_{2}^{\pm}K_{\Sigma}(R_{\Sigma}K_{\Sigma})^{-1}R_{\Sigma} c_{2}^{\mp}$ is smoothing, or using Lemma \ref{blip} that $ c_{1}^{\pm}(R_{\Sigma}K_{\Sigma})^{-1}R_{\Sigma} c_{2}^{\mp}$ is smoothing. This follows from Prop. \ref{prop10.1} (2)  and (3). \qed

\begin{remark}
 Property (4) in Prop. \ref{prop10.2} is the key property of the microlocal  TT-synchronous gauge condition. It will be used in the next subsection for the construction of Hadamard states. It is not satisfied by the usual TT-synchronous gauge condition.
\end{remark}

\subsection{Gauge fixing in the singular case}\label{sec10.5}
Assume now that $R_{\Sigma}K_{\Sigma}$ is {\em not invertible}. 

\subsubsection{Notation}

The charges   $q_{2}$ and hence $q_{I, 2}$ are well defined on $\cH^{s}$ for $s\geq \12$, which we will assume in the sequel. The orthogonal of a subspace $E\subset \cH^{s}$ for $q_{I, 2}$ will be denoted by $E^{q_{I, 2}}$. 

%

If $A: \cinf(\Sigma; \tV_{k}\otimes\cc^{2})\to \cinf(\Sigma; \tV_{1}\otimes \cc^{2})$ we define $A^{\dag}: \cD'(\Sigma; \tV_{1}\otimes \cc^{2})\to \cD'(\Sigma; \tV_{k}\otimes \cc^{2})$ by
\[
(u| A f)_{L^{2}(\Sigma; \tV_{1}\otimes \cc^{2})}\defeq \bar{A^{\dag}u}\dual q_{I, k}f,
\]
where we recall that $q_{I, 2}= q_{2}\circ I$, $q_{I, 1}= q_{1}$. 
Denoting by $A^{*}$ the usual adjoint obtained from the Hilbertian structure of $\tV_{k}\otimes \cc^{2}$ we have $A^{*}= q_{I, k}A^{\dag}$.

\subsubsection{The space $\cH^{s}_{\rm reg}$}
We set:
\[
\cH^{s}_{\rm reg}(\Sigma; \tilde{V}_{2}\otimes \cc^{2})\defeq \{f_{2}\in \cH^{s}(\Sigma; \tilde{V}_{2}\otimes \cc^{2}): R_{\Sigma}f_{2}\in \Ran R_{\Sigma}K_{\Sigma}\},
\]
and we will sometimes  write $\cH^{s}_{{\rm reg}}$ for  $\cH^{s}_{{\rm reg}}(\Sigma; \tilde{V}_{2}\otimes \cc^{2})$.  The gauge fixing equation \eqref{e10.-4b} can be solved  iif $f_{2}\in \cH^{s}_{\rm reg}$. Clearly
\[
\Ker R_{\Sigma}\cap \cH^{s}\subset \cH^{s}_{\rm reg}, \ K_{\Sigma}\cH^{s+1}\subset \cH^{s}_{\rm reg}.
\]

By ellipticity $\Ker (R_{\Sigma}K_{\Sigma})^{*}$  is a finite dimensional space in $ \cinf(\Sigma; \tilde{V_{1}}\otimes \cc^{2})$,  and
 there exist $u_{i}\in \Ker (R_{\Sigma}K_{\Sigma})^{*}$, $1\leq i\leq n$ such that $u\in \Ran R_{\Sigma}K_{\Sigma}$ iff
$(u_{i}| u)_{\tV_{1}\otimes \cc^{2}}=0$, $1\leq i\leq n$.

A routine computation gives that
\[
R_{\Sigma}^{\dag}= \mat{(b^{+}-b^{-})^{-1}}{(b^{+}-b^{-})^{-1}}{b^{+}(b^{+}-b^{-})^{-1}}{-b^{-}(b^{+}-b^{-})^{-1}}\times \mat{J}{0}{0}{J},
\]
for $J:H^{s}(\Sigma; \tilde{V}_{1})\to H^{s}(\Sigma; \tilde{V}_{2})$ defined by
\[
(Jv)_{tt}= \12 v_{t}, \ (Jv)_{t\Sigma}= v_{\Sigma}, \ (Jv)_{\Sigma\Sigma}= - \12 v_{t}\rh_{0}.
\]
In particular we see that $R_{\Sigma}^{\dag}: \cinf(\Sigma; \tilde{V}_{1}\otimes \cc^{2})\to \cinf(\Sigma; \tilde{V}_{2}\otimes \cc^{2})$,
hence  $v_{i}= R_{\Sigma}^{\dag}u_{i}\in \cinf(\Sigma; \tilde{V}_{2}\otimes \cc^{2})$. Since $(R_{\Sigma}K_{\Sigma})^{*}u_{i}=q_{1}K_{\Sigma}^{\dag}R_{\Sigma}^{\dag}u_{i}=0$ we get
\[
K_{\Sigma}^{\dag}v_{i}=0.
\]
Without loss of generality, we can assume that the $v_{i}$ are linearily independent and we set $\mathcal{V}= {\rm Vect}\{v_{i}: 1\leq i\leq n\}$. 
We have
\[
\cH^{s}_{\rm reg}= \{f_{2}\in \cH^{s}: \bar{v}_{i}\dual q_{I, 2}f_{2}=0, \ 1\leq i\leq n\}= \cV^{q_{I,2}},
\]
and $\cH^{s}_{\rm reg}$ is of codimension $n$ in $\cH^{s}$. 
Let us pick a supplementary space $\cV_{1}$ in $\cV$ of $\cV\cap \Ran K_{\Sigma}$.  We can assume that $v_{1}, \dots, v_{p}\in \cV_{1}$ and $v_{p+1}, \dots, v_{n}\in \cV\cap \Ran K_{\Sigma}$. Since $\Ker K_{\Sigma}^{\dag}= \Ran K_{\Sigma}^{q_{I,2}}$, we obtain that
\[
\Ker K_{\Sigma}^{\dag}\cap \cH^{s}_{\rm reg}= \Ker K_{\Sigma}^{\dag}\cap \cV_{1}^{q_{I,2}},
\]
and $\Ker K_{\Sigma}^{\dag}\cap \cH^{s}_{\rm reg}$ is of codimension $p$ in $\Ker K_{\Sigma}^{\dag}$.  Therefore we can find $w_{1}, \dots, w_{p}\in \Ker K_{\Sigma}^{\dag}\cap \cinf(\Sigma; \tV_{2}\otimes \cc^{2})$ such that $A= [\bar{w}_{i}\dual q_{I, 2}v_{j}]_{1\leq i, j\leq p}$ is invertible. 

Since $\cH^{s}_{\rm reg}$ is of codimension $n$ in $\cH^{s}$, we can   complete the $w_{i}$ for $1\leq i\leq p$ by $w_{i}\in \cinf$ for $p+1\leq i\leq n$ so that $Q= [\bar{w}_{i}\dual q_{2}v_{j}]_{1\leq i,j\leq n}$ is invertible.   Using that $w_{i}\in \Ker K_{\Sigma}^{\dag}$ and $v_{j}\in \Ran K_{\Sigma}$ for $i\leq p, j\geq p+1$ we have $Q= \mat{A}{0}{B}{C}$, where $A\in M_{p}(\cc)$, $C\in M_{n-p}(\cc)$ are invertible.
Replacing $w_{i}$ by $\sum_{k}\bar{T}_{ik}w_{k}$ for $T= \mat{A^{-1}}{0}{-C^{-1}B A^{-1}}{C^{-1}}$ we obtain
\beq\label{e10.7b}
\begin{array}{l}
\bar{w}_{i}\dual q_{I, 2}v_{j}= \delta_{ij},\\[2mm]
 w_{i}\in \Ker K_{\Sigma}^{\dag}\hbox{ for } i\leq p, \\[2mm]
 v_{j}\in \Ker K_{\Sigma}^{\dag}\hbox{ for }j\leq n, \\[2mm]
 v_{j}\in \Ran K_{\Sigma}\hbox{ for }j\geq p+1.
\end{array}
\eeq
We set now
\begin{equation}
\label{e10.8}
\pi f_{2}= \sum_{j=1}^{n}\bar{v}_{j}\dual q_{I, 2}f_{2} w_{j}.
\end{equation}

\begin{lemma}\label{lemm10.2}
\ben
\item $1- \pi: \cH^{s}\to \cH^{s}_{\rm reg}$ is a projection;
\item $\pi\circ K_{\Sigma}=0$;
\item $1- \pi$ preserves  $\Ker K_{\Sigma}^{\dag}$;
\item $\pi^{\dag}$ preserves $\Ran K_{\Sigma}$.
\een
\end{lemma}
\proof The fact that $\pi$ is a projection is easy. We have $\Ker \pi= \cH^{s}_{\rm reg}$ which proves (1).   We have $\pi\circ K_{\Sigma}=0$ since $v_{i}\in \Ker K_{\Sigma}^{\dag}$ and  \[
\pi^{\dag}f_{2}= \sum_{j=1}^{n}\bar{w}_{j}\dual q_{I, 2}f_{2}v_{j}.
\]
By \eqref{e10.7b} this implies that $\pi^{\dag}$ preserves $\Ran K_{\Sigma}$ which implies that $\pi$ preserves $\Ker K_{\Sigma}^{\dag}$. \qed

Let us now construct the analog of the projection $T$ in the singular case. We fix two projections
\[
\begin{array}{rl}
\pi_{1}: \cH^{s}(\Sigma; \tV_{1}\otimes \cc^{2})\to \Ker R_{\Sigma}K_{\Sigma},\\[2mm]
\pi_{2}: \cH^{s}(\Sigma; \tV_{2}\otimes \cc^{2})\to  \Ran K_{\Sigma}\cap \Ker R_{\Sigma}.
\end{array}
\]
The projections $\pi_{i}$ are finite rank with smooth distributional kernels. Let us define the map
\[
(R_{\Sigma}K_{\Sigma})^{-1}R_{\Sigma}: \cH^{s}_{\rm reg}(\Sigma; \tV_{2}\otimes \cc^{2})\to \cH^{s+1}(\Sigma; \tV_{1}\otimes \cc^{2})\cap \Ker \pi_{1},
\]
such that  $f_{1}= (R_{\Sigma}K_{\Sigma})^{-1}R_{\Sigma}f_{2}$ for $f_{2}\in \cH^{s}_{\rm reg}$ is the unique solution of
\beq\label{e-06}
\left\{
\begin{array}{l}
R_{\Sigma}K_{\Sigma}f_{1}= R_{\Sigma}f_{2}, \\[2mm]
 \pi_{1}f_{1}=0.
\end{array}\right.
\eeq
We set 
\beq\label{e10.20}
T_{\rm reg}=\one - K_{\Sigma}(R_{\Sigma}K_{\Sigma})^{-1}R_{\Sigma}: \cH^{s}_{\rm reg}(\Sigma; \tV_{2}\otimes\cc^{2})\to  \cH^{s}(\Sigma; \tV_{2}\otimes\cc^{2}).
\eeq

\begin{lemma}\label{lemm10.4}
  \ben
 \item \[
 T_{\rm reg}:\cH^{s}_{\rm reg}(\Sigma; \tV_{2}\otimes \cc^{2})\to\Ker R_{\Sigma}
  \]
  is a projection;
  \item $  T_{\rm reg}$ preserves $\cH^{s}_{\rm reg}(\Sigma; \tV_{2}\otimes \cc^{2})\cap \Ker K_{\Sigma}^{\dag}$;
  \item $  T_{\rm reg}K_{\Sigma}= K_{\Sigma}\pi_{1}$.
  \een

\end{lemma}
 \proof We  have $R_{\Sigma}T_{\rm reg}=0$ hence $\Ran T_{\rm reg}\subset \Ker R_{\Sigma}$ and $ T_{\rm reg}= \one $ on $\Ker R_{\Sigma}$ which implies (1).   (2) follows from $K_{\Sigma}^{\dag}K_{\Sigma}=0$. If $f_{2}= K_{\Sigma}g_{1}$, the unique solution of \eqref{e-06} is $f_{1}= (1-\pi_{1})g_{1}$  which implies (3). \qed
 
 We set now
\beq\label{e10.21}
T= (1-\pi_{2})\circ T_{\rm reg}\circ (1-\pi):\cH^{s}(\Sigma; \tV_{2}\otimes \cc^{2})\to \cH^{s}(\Sigma; \tV_{2}\otimes \cc^{2}).
\eeq
\begin{proposition}\label{prop10.3}
 \ben
 
 \item $T: \cH^{s}\to  \Ker R_{\Sigma}\cap \Ker \pi_{2}$ is a projection;
  \item $TK_{\Sigma}=0$;
  \item $T$ preserves $\Ker K_{\Sigma}^{\dag}$;
  \item $c_{2}^{\pm}T c_{2}^{\mp}$ is smoothing.
 \een
\end{proposition}
\proof  
 From Lemma \ref{lemm10.4} (1) and $\Ran \pi_{2}\subset \Ker R_{\Sigma}$ we get that  $R_{\Sigma}T=0$, so $\Ran T\subset \Ker R_{\Sigma}\cap \Ker \pi_{2}$. Since $\Ker R_{\Sigma}\subset \cH^{s}_{\rm reg}$ we have  $\pi=0$ on $\Ker R_{\Sigma}$, hence $ T_{\rm reg}(1-\pi)=\one$ on $\Ker R_{\Sigma}$ hence $T= \one$ on $\Ker R_{\Sigma}\cap \Ker \pi_{2}$. This proves (1).

By Lemma \ref{lemm10.2} (2) we have $(1- \pi)K_{\Sigma}= K_{\Sigma}$ hence $TK_{\Sigma}= (1- \pi_{2})K_{\Sigma}\pi_{1}=0$ by Lemma \ref{lemm10.4} (3) and the definition of $\pi_{1}, \pi_{2}$. This proves (2).

(3) follows from Lemma \ref{lemm10.4} (2) and Lemma \ref{lemm10.2} (3). To prove (4) we argue  as in the proof of Prop. \ref{prop10.2} (3) using  Prop. \ref{prop10.1}, and additionally the fact that $\pi, \pi_{2}$ have smooth distributional kernels.  \qed

\section{Hadamard states}\init\label{sec11}
In this section we construct a Hadamard state by modifying the Hadamard projectors $c_{2}^{\pm}$ using the projection $T$. As in the previous section we assume that $\Sigma$ is compact.

We start with the simpler regular case.
\subsection{Regular case}
Let us set
\beq\label{e11.9}
\lambda_{2\Sigma}^{\pm}\defeq \pm T^{*}\circ q_{I, 2}c_{2}^{\pm}\circ T.
\eeq
and as in \ref{sec1.7.3}:
\[
\Lambda_{2}^{\pm}\defeq (\rho_{2}G_{2})^{*}\lambda_{2\Sigma}^{\pm}(\rho_{2}G_{2}).
\]

\begin{theoreme}\label{thm11.1}
 The pair $\lambda_{2\Sigma}^{\pm}$ are the Cauchy surface covariances of a gauge invariant Hadamard state for $P$, i.e.
\ben
\item  $\WF(\Lambda_{2}^{\pm})'\subset \cN^{\pm}\otimes \cN^{\pm}$.
\item $\lambda_{2\Sigma}^{+}- \lambda_{2\Sigma}^{-}= q_{I, 2}$ on $\Ker K_{\Sigma}^{\dag}$,
\item $\lambda_{2\Sigma}^{\pm}=0$ on $\Ker_{\c}K_{\Sigma}^{\dag}\times \Ran K_{\Sigma}$
\item $\lambda_{2\Sigma}^{\pm}= \lambda_{2\Sigma}^{\pm*}$, $\lambda_{2\Sigma}^{\pm}\geq 0$ on $\Ker_{\c} K_{\Sigma}^{\dag}$.
\een
Therefore $\Lambda_{2}^{\pm}$ are the covariances of a quasi-free Hadamard state on $\CCR(\cV_{P}, Q_{P})$.
\end{theoreme}
\proof 
Let us first prove (1). We apply \cite[Prop 11.1.1]{G}. Let $U_{2}(t, s)$ be the Cauchy evolution for  $D_{2}$. Writing $T^{*}q_{I,2}\eqdef q_{I, 2}T^{\dag}$, we  need to show that $\WF(U_{2}(\cdot, 0))T^{\dag} c_{2}^{\pm}T)'\subset (\cN^{\pm}\cup \cF)\times T^{*}\Sigma$ for some conic set $\cF\subset T^{*}M$ with $\cF \cap \cN= \emptyset$.

By Prop. \ref{prop10.2} $ c_{2}^{\pm}T  c_{2}^{\mp}$ is smoothing, hence $ c_{2}^{\pm}T^{\dag}  c_{2}^{\mp}$ is smoothing, since $ c_{2}^{\pm}= ( c_{2}^{\pm})^{\dag}$.  Therefore $U_{2}(\cdot, 0)T^{\dag} c_{2}^{\pm}T= U_{2}(\cdot, 0) c_{2}^{\pm}T^{\dag}T$ modulo a smooth kernel.  So the result follows from the Hadamard property of $ c_{2}^{\pm}$.

Let us now prove (2).   We have $\lambda_{2}^{+}- \lambda_{2}^{-}= T^{*}q_{I, 2}T$. 
$T$ preserves $\Ker K_{\Sigma}^{\dag}$ by Prop. \ref{prop10.2} (3) and $Tf_{2}= f_{2}$ mod $\Ran K_{\Sigma}$ hence $\lambda_{2}^{+}- \lambda_{2}^{-}= q_{I, 2}$ on $\Ker K_{\Sigma}^{\dag}$.

(3) follows from Prop. \ref{prop10.2} (2).

Since  $\Ran T\subset \Ker R_{\Sigma}$ and $\pm q_{2}c_{2}^{\pm}\geq 0$ on $\Ker R_{\Sigma}$ by Lemma \ref{lemma.e-02}, we obtain (4). 
 \qed

\subsection{Singular case} We now consider the singular case.  We need an additional modification  of $c_{2}^{\pm}$ since  $T^{*}q_{I, 2}T\neq q_{I, 2}$ on $\Ker K_{\Sigma}^{\dag}$ because of the projection $\pi$.  This modification is inspired by  a construction in \cite[Subsect. 4.4]{FS}.
Let
\[
\tilde{\pi}f_{2}= \sum_{j=1}^{p}\bar{v}_{j}\dual q_{I, 2}f_{2}w_{j},
\]
and note using \eqref{e10.7b} that $\tilde{\pi}= \pi$ on $\Ker K_{\Sigma}^{\dag}$.

\def\tpi{\tilde{\pi}}
Consider the hermitian form \beq\label{e11.12}
\begin{array}{rl}
\nu\defeq &q_{I, 2}- (1-\tpi)^{*}q_{I, 2}(1- \tpi)\\[2mm]
=&q_{I, 2}\tpi^{\dag}+ q_{I, 2} \tpi- q_{2}\tpi^{\dag}\tpi.
\end{array}
\eeq
acting on $\cinf(\Sigma; \tV_{2}\otimes \cc^{2})$. It has a smooth distributional kernel and a finite rank. Identifying it with a selfadjoint  operator using the Hilbertian scalar product $(\cdot| \cdot)_{\tV_{2}\otimes \cc^{2}}$, we can find $u_{i}\in \cinf(\Sigma; \tV_{2}\otimes \cc^{2})$, $1\leq i\leq q$ linearily independent such that 
\[
\nu= \sum_{i=1}^{q}\alpha_{i}|u_{i})(u_{i}|, \ \alpha_{i}\neq 0.
\]
We set
\beq\label{e11.10}
\bar{f}_{2}\dual \lambda_{2\Sigma}^{\pm} f_{2}\defeq\pm \bar{Tf_{2}}\dual q_{I, 2}c_{2}^{\pm}Tf_{2} \pm (f_{2}| \one_{\rr^{\pm}}(\nu)\nu f_{2})_{\tV_{2}\otimes \cc^{2}},
\eeq
\[
\Lambda_{2}^{\pm}\defeq (\rho_{2}G_{2})^{*}\lambda_{2\Sigma}^{\pm}(\rho_{2}G_{2}).
\]

\begin{theoreme}\label{thm11.2}
 The pair $\lambda_{2\Sigma}^{\pm}$ are the Cauchy surface covariances of a gauge invariant Hadamard state for $P$, i.e.
\ben
\item  $\WF(\Lambda_{2}^{\pm})'\subset \cN^{\pm}\otimes \cN^{\pm}$.
\item $\lambda_{2\Sigma}^{+}- \lambda_{2\Sigma}^{-}= q_{I, 2}$ on $\Ker K_{\Sigma}^{\dag}$,
\item $\lambda_{2\Sigma}^{\pm}=0$ on $\Ker_{\c}K_{\Sigma}^{\dag}\times \Ran K_{\Sigma}$
\item $\lambda_{2\Sigma}^{\pm}= \lambda_{2\Sigma}^{\pm*}$, $\lambda_{2\Sigma}^{\pm}\geq 0$ on $\Ker_{\c} K_{\Sigma}^{\dag}$.
\een
Therefore $\Lambda_{2}^{\pm}$ are the covariances of a quasi-free Hadamard state on $\CCR(\cV_{P}, Q_{P})$.\end{theoreme}
\proof 
The proof of (1) is identical to Thm. \ref{thm11.1}. Note that the additional term in \eqref{e11.10} produces a smooth additional term in  the two-point functions $\Lambda_{2}^{\pm}$.

Let us prove (2).  We have using \eqref{e11.12}:
\[
\lambda_{2\Sigma}^{+}- \lambda_{2\Sigma}^{-}= T^{*}q_{I, 2}T+ q_{I, 2}- (1- \tpi)^{*}q_{I, 2}(1- \tpi).
\]
Let us compute the first term in the r.h.s. on $\Ker K_{\Sigma}^{\dag}$. Recall that $T= (1- \pi_{2})T_{\reg}(1- \pi)$. We   have $T_{\rm reg}(1- \pi)f_{2}= (1- \pi)f_{2}$ modulo  $\Ran K_{\Sigma}$ by \eqref{e10.20}. Since $\Ran \pi_{2}\subset\Ran K_{\Sigma}$ we obtain that $Tf_{2}= (1- \pi)f_{2}$ modulo  $\Ran K_{\Sigma}$. By Prop. \ref{prop10.3} $T$ preserves $\Ker K_{\Sigma}^{\dag}$ hence:
\[
 \bar{Tf_{2}}\dual q_{I, 2}Tf_{2}= \bar{(1-\pi)f_{2}}\dual q_{I,2}(1- \pi)f_{2}
 =\bar{(1-\tpi)f_{2}}\dual q_{I, 2}(1- \tpi)f_{2}, 
\]
since $\pi= \tilde{\pi}$ on  $\Ker K_{\Sigma}^{\dag}$.
Using the definition of $\nu$ in \eqref{e11.12}  we obtain that $\lambda_{2\Sigma}^{+}- \lambda_{2\Sigma}^{-}= q_{I, 2}$ on $\Ker K_{\Sigma}^{\dag}$.

We now prove (3). By Prop. \ref{prop10.3} (2) $TK_{\Sigma}=0$. Since $v_{j}, w_{j}\in \Ker K_{\Sigma}^{\dag}$ for $1\leq j\leq p$, see \eqref{e10.7b}, we obtain that $\tilde{\pi}K_{\Sigma}= \tilde{\pi}^{\dag}K_{\Sigma}=0$.  Therefore $\nu=0$ on $\Ran K_{\Sigma}$ hence $\lambda_{2\Sigma}^{\pm}= 0$ on $\Ran K_{\Sigma}$. This proves (3). 
  
It remains to prove (4). The first term in the rhs of \eqref{e11.10} is positive  by the same argument as in the proof of Thm. \ref{thm11.2}, since $\Ran T\subset \Ker R_{\Sigma}$. The second term is also clearly positive. This completes the proof of the theorem. \qed


\begin{thebibliography}{aaa}
\bibitem[AA]{AA}Ashtekar A., Magnon-Ashtekar, A.: On the symplectic structure of general relativity, Comm. Math. Phys. {\bf 86} (1982), 55-68.
\bibitem[BDM]{BDM}Benini M.,  Dappiagi C.,  Murro S.: Radiative observables for linearized gravity on asymptotically flat spacetimes and their boundary induced states, J. Math. Phys. {\bf 55} (2014), 082301.
\bibitem[B]{B}Boucetta M.: Spectre des Laplaciens de Lichnerowicz sur les sphères et les projectifs réels, Publications Math. {\bf 43} (1999), 451-483.
\bibitem[BFR]{BFR}Brunetti R., Fredenhagen K., Rejzner K.: Quantum gravity from the point of view of locally covariant quantum field theory, Comm. Math. Phys. {\bf 345} (2016), 741-779.
\bibitem[DMP]{DMP}Dappiagi C., Moretti W., Pinamonti N.: {\em Hadamard states from light-like hypersurfaces}  Springer Briefs in Mathematical Physics {\bf 25} (2017).
\bibitem[DS]{DS}Dappiagi C., Siemssen D.: Hadamard states for the vector potential on asymptotically flat spacetimes, Rev. Math. Phys. {\bf 25} (2013) 1350002.
\bibitem[FH]{FH}Fewster C., Hunt D.: Quantization of linearized gravity in cosmological vacuum spacetimes, Rev. Math. Phys. {\bf 25} (2013), 1330003.
\bibitem[FP]{FP}Fewster C., Pfenning, M.: A quantum weak energy inequality for spin-one fields in curved space–time, J. Math. Phys. {\bf 44} (2003) 4480-4079.
\bibitem[FS]{FS}Finster F., Strohmaier A.: Gupta–Bleuler quantization of the Maxwell field in globally hyperbolic spacetimes, Ann. Henri Poincaré {\bf 16} (2015), 1837-1868.
\bibitem[F]{F}Furlani E.: Quantization of the electromagnetic field on static space–times, J. Math. Phys. {\bf 36} (1995), 1063-1079.
 \bibitem[GMW]{GMW} Gérard C., Murro S., Wrochna M.: Quantization of linearized gravity by Wick rotation in Gaussian time, (2022) ArXiv preprint 2204.01094.
 \bibitem[GOW]{GOW} G\'erard C., Oulghazi 0., Wrochna M.: Hadamard states for the Klein-Gordon equation on Lorentzian manifolds of bounded geometry, Comm. Math. Phys. {\bf 352} (2017), 519-583.
\bibitem[GW1]{GW1} G\'erard C., Wrochna M.: Construction of Hadamard states by pseudodifferential calculus, Comm. Math. Phys. {\bf 325}  (2014), 713-755.
\bibitem[GW2]{GW2} G\'erard C., Wrochna M.: Hadamard states for the linearized Yang–Mills equation on curved spacetime, Comm. Math. Phys. {\bf 337} (2015), 253-320.
\bibitem[GW3]{GW3} G\'erard C., Wrochna M.: Analytic Hadamard states, Calder{\'{o}}n projectors and Wick rotation near analytic Cauchy surfaces, Comm. Math. Phys. {\bf 366} (1019), 29-65.
\bibitem[G]{G}Gérard C.: {\em Microlocal Analysis of Quantum Fields on Curved Spacetimes}, ESI Lectures in Mathematics and Physics EMS (2019).
\bibitem[HS]{HS}Hack T., Schenkel A.: Linear bosonic and fermionic quantum gauge theories on curved spacetimes Gen. Relativ. Grav. {\bf 45} (2013), 877-910.
\bibitem[H]{H}Hollands S.: Renormalized quantum Yang-Mills fields in curved spacetime, Rev. Math. Phys. {\bf 20} (2008), 1033-1172.
  \bibitem[J1]{junker}Junker W.: {\em Adiabatic Vacua and Hadamard States for Scalar Quantum Fields on Curved Space-time}, DESY-thesis-1995-144, (1995). ArXiv preprint hep-th/9507097v1.
  \bibitem[J2]{junkererratum}Junker W.: Erratum to "Adiabatic Vacua and Hadamard States for Scalar Quantum Fields on Curved Space-time", Rev. Math. Phys. {\bf 207} (2002), 511-517.  
 \bibitem[L]{L}Lichnerowicz A.: Propagateurs et commutateurs on relativité générale, Publ. Math.  I.H.E.S.  {\bf 10} (1961), 5-56.
 \bibitem[R]{R}Ringstr\"{o}m H. : {\em The Cauchy Problem in General Relativity} ESI Lectures in Mathematics and Physics EMS (2009).
 \bibitem[SV]{SV}Sahlmann H.,   Verch R.:  Microlocal spectrum condition and Hadamard form for vector-valued
quantum fields in curved spacetime. Rev. Math. Phys., {\bf 13} (2001), 1203–1246.

  
  \end{thebibliography}
\end{document}